\documentclass[aps, prl, preprint, superscriptaddress, showpacs]{revtex4-1}
\usepackage{graphics}
\usepackage[english]{babel}
\usepackage{epstopdf}
\usepackage{amsmath}
\usepackage{amsfonts}
\usepackage{amssymb}
\usepackage[latin1]{inputenc} %caratteri speciali
\usepackage[OT2,T1]{fontenc}

\begin{document}

%\preprint{}



%%%% AUTHOR LIST

\author{Simone Finizio}
\email{simone.finizio@psi.ch}
\affiliation{Paul Scherrer Institut, 5232 Villigen PSI, Switzerland}

\author{Claire Donnelly}
\affiliation{Cavendish Laboratory, University of Cambridge, Cambridge CB3 0HE, United Kingdom}
\affiliation{Max Planck Institute for Chemical Physics of Solids, 01187 Dresden, Germany}

\author{Sina Mayr}
\affiliation{Paul Scherrer Institut, 5232 Villigen PSI, Switzerland}
\affiliation{Laboratory for Mesoscopic Systems, Department of Materials, ETH Zurich, 8093 Zurich, Switzerland}

\author{Ales Hrabec}
\affiliation{Paul Scherrer Institut, 5232 Villigen PSI, Switzerland}
\affiliation{Laboratory for Mesoscopic Systems, Department of Materials, ETH Zurich, 8093 Zurich, Switzerland}

\author{J\"org Raabe}
\affiliation{Paul Scherrer Institut, 5232 Villigen PSI, Switzerland}

%%%% END OF AUTHOR LIST

\title{Three-dimensional Resonant Magnetization Dynamics Unraveled by Time-Resolved Soft X-ray Laminography}
%\keywords{magnetic vortex gyration, 3D time-resolved magnetic imaging, scanning transmission X-ray microscopy, laminography, magnetization dynamics}

\date{\today}

%\doublespacing

%\begin{document}

\begin{abstract}
The imaging of magneto-dynamical processes has been, so far, mostly a two-dimensional business, due to the constraints of the available experimental techniques. In this manuscript, building on the recent developments of soft X-ray magnetic laminography, we present an experimental setup where magneto-dynamical processes can be resolved in all three spatial dimensions and in time, with the possibility to freely tune the frequency of the dynamical process. We then employ this setup to investigate the three-dimensional dynamics of two resonant magneto-dynamical modes in a CoFeB microstructure occurring at different frequencies, namely the fundamental vortex gyration mode and a magnetic field-induced domain wall excitation mode. This new technique provides much needed capabilities for the experimental investigation of the magnetization dynamics of three-dimensional magnetic systems.
\end{abstract}


\maketitle

%-- Introduction

%- Magnetization dynamics

Magneto-dynamical processes encompass a wide spectrum of phenomena, including vortex dynamics \cite{art:guslienko_gyrovector, art:guslienko_vortexMass, art:krueger_harmonicPotential}, magnonics \cite{art:dieterle_SpinWaves, art:wintz_spin_waves, art:mayr_spinwaves, art:divinskiy_BEC_SOT}, current and field induced domain wall motion \cite{art:jue_FIDWM, art:rhensius_field_domain_motion, art:ryu_CIDWM}, skyrmion dynamics \cite{art:kai_skyrmion_hall_angle, art:woo_skyrmion_nucleation, art:finizio_SkyrmionNucleation, art:buettner_gyration}, and magnetization switching \cite{art:baumgartner_switching}. The understanding of magneto-dynamical processes is vital not only for the design and operation of magnetic devices, but also for the confirmation of theoretical models. Experimental techniques such as time-resolved X-ray magnetic microscopy have provided vital contributions towards the unraveling of magneto-dynamical processes at the nanoscale.

%- Extension to three dimensions

Almost all of the magneto-dynamical processes that have been imaged have been on two-dimensional systems. The extension of such investigations onto the third spatial dimension is of great interest both fundamentally and for applications, thanks to the rich dynamical behavior offered by three-dimensional spin configurations \cite{art:fernandez_3DMagnetism, art:fisher_3DMagnetism}. Examples of such three-dimensional dynamical modes include higher order spin wave processes \cite{art:dieterle_SpinWaves} and their generation from three-dimensional topological features \cite{art:mayr_spinwaves}. However, until recently, the only experimental investigations that could be carried out on three-dimensional systems have been limited to indirect characterizations, performed through the imaging of two-dimensional projections followed by comparisons with theoretical predictions and micromagnetic simulations \cite{art:wartelle_bloch_points, art:mayr_spinwaves}.

In recent years, significant progress has been made towards the imaging of magneto-dynamical processes resolved in all three spatial dimensions, driven by the development of three-dimensional X-ray imaging techniques such as magnetic tomography \cite{art:donnelly_3DImaging} and laminography \cite{art:donnelly_TRLamni, art:witte_Lamni}. Of particular interest is the work reported in Ref. \cite{art:donnelly_TRLamni}, where the dynamics of a 1.2 $\mu$m thick GdCo microstructured element were, for the first time, resolved in all three dimensions. However, this groundbreaking study relied on the intrinsically pulsed structure of synchrotron radiation, investigating a dynamical process locked to the pulse repetition rate of the synchrotron light source (500 MHz in the case of Ref. \cite{art:donnelly_TRLamni}) to perform the time-resolved imaging. This is caused by the requirement to utilize a two-dimensional X-ray detector demanded by the ptychographic imaging technique used to acquire each projection \cite{art:holler_lamni}. The limited bandwidth of two-dimensional detectors allows only for the imaging of processes occurring at either the master clock frequency or its harmonics. This approach therefore limits the accessible frequencies, directly impacting the ensemble of dynamical processes that can be investigated, as many such processes are resonant at particular frequencies. As a result, to fully probe magnetization dynamics in three dimensions, the free control over the frequency of the dynamical process is key. In this work, we achieve frequency-flexible three-dimensional magnetic imaging through the combination of soft X-ray magnetic laminography with time-resolved scanning transmission X-ray microscopy (STXM) imaging without constraints on the excitation frequency, and exploiting the stronger XMCD contrast of magnetic elements at soft X-ray energies.

%-- Experiments

%- Description of TR-STXM + lamni imaging

\begin{figure}[p]
 \includegraphics{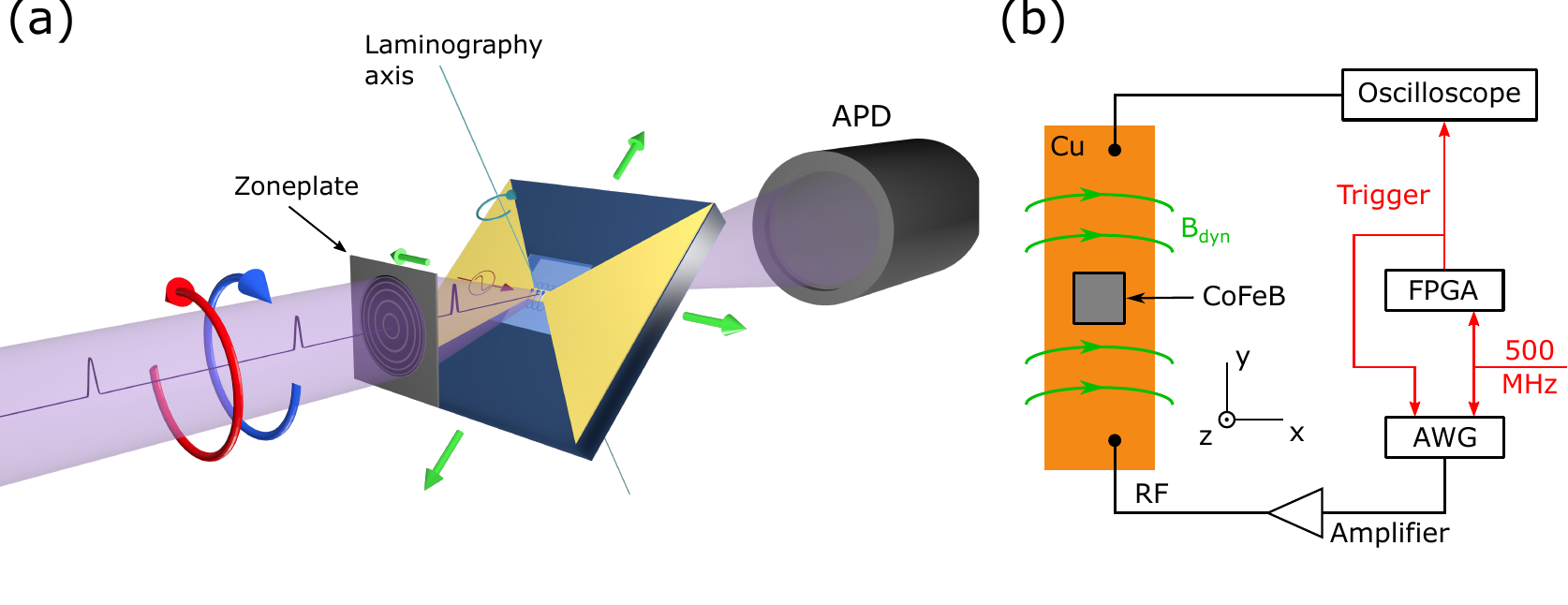}
  \caption{(a) Sketch depicting the geometry and coordinate system for the time-resolved laminography experiments presented in this work. The sample is mounted under an angle of 45$^\circ$ with respect to the X-ray beam. To acquire the projections necessary for the laminographic reconstruction, the sample is rotated at different angles around the axis perpendicular to its surface ($z$ axis). (b) Sketch of the electrical configuration utilized for the generation of the oscillating magnetic field used to excite the dynamical processes presented here. An arbitrary waveform generator (AWG), frequency locked to the 500 MHz master clock of the synchrotron light source, is used to inject a RF current across a stripline fabricated above a Co$_{40}$Fe$_{40}$B$_{20}$ microstructured square, giving rise to an oscillating magnetic field along the $x$ axis through the Oersted effect. The red signals indicate synchronization and timing signals (handled by a custom designed field programmable gate array - FPGA - setup).}
 \label{fig:lamni_setup}
\end{figure}

The three-dimensional imaging of thin film samples exhibits some unique challenges compared to the standard cylindrical geometries utilized for tomographic imaging. These additional challenges are caused by the tomography imaging geometry, where the tomographic rotation axis is perpendicular to the imaging beam axis, i.e. lies in the surface plane of the thin film sample. For projections at angles close to 90$^\circ$, the large apparent thickness of the thin film sample hinders the acquisition of the projection, leading to a missing wedge artifact in the Fourier space. Furthermore, the change in apparent thickness of the thin film sample as a function of the tomographic rotation angle has to be accounted for in the reconstruction. To overcome these two issues, laminographic imaging has been developed. Laminography is a generalization of the tomographic imaging protocol where the sample rotation axis is not perpendicular to the imaging beam axis, but is rather oriented at a defined angle (in our case, 45$^\circ$) with respect to the beam axis, as sketched in Fig. \ref{fig:lamni_setup}(a) \cite{art:holler_lamni, art:witte_Lamni, art:donnelly_TRLamni}. The laminography geometry probes a larger volume of the Fourier space (missing cone artifact \cite{art:holler_lamni}), which allows for an improvement of the quality of the three-dimensional reconstruction. Furthermore, the rotation geometry is such that no changes in the effective thickness of the sample occur when rotating it about the laminography axis, simplifying the measurements \cite{art:holler_lamni, art:witte_Lamni, art:donnelly_TRLamni}. Finally, laminography allows for the imaging of extended sample surfaces, as demonstrated in Ref. \cite{art:holler_lamni}.

%\section{Results and Discussion}

To be able to visualize the magnetization dynamics of three-dimensional systems, the development of a technique that is able to perform time-resolved imaging with freely selectable frequencies combined with three-dimensional imaging at soft X-ray energies is therefore crucial. In this work, we have combined time-resolved imaging with the soft X-ray laminography setup developed at the PolLux endstation of the Swiss Light Source \cite{art:witte_Lamni}, where each laminography projection is acquired through STXM imaging. Compared to the ptychographic imaging used in Ref. \cite{art:donnelly_TRLamni}, STXM has the advantage of being a scanning microscopy technique requiring a point detector. By using an avalanche photodiode (APD) with a bandwidth larger than the repetition rate of the synchrotron light source (for the Swiss Light Source, 500 MHz), two consecutive X-ray pulses generated by the light source (separated by 2 ns) can be resolved by the APD. This allows for the possibility to sort each recorded photon count by the APD into separate time channels depending on the relative phase between when the count occurred and the excitation signal used to trigger the dynamical process, which is synchronized to a rational multiple of the repetition rate \cite{art:puzic_TR_STXM}. This is achieved through a custom-designed field-programmable gate array (FPGA) setup. Thanks to this detection protocol, a larger comb of accessible frequencies, given by rational multiples of the master clock frequency, is available. Purely arbitrary excitation frequencies are still not accessible, but this final limitation can be easily lifted if time-of-arrival detection is performed \cite{art:finizio_QuTAG}. For the work presented in this manuscript, the time-resolved imaging was performed using the FPGA setup. Further details about the time-resolved imaging setup are given in the Methods section.

%- Measurements as a function of the frequency - VC gyration and SW in DW mode

%% Statics

\begin{figure}[p]
 \includegraphics{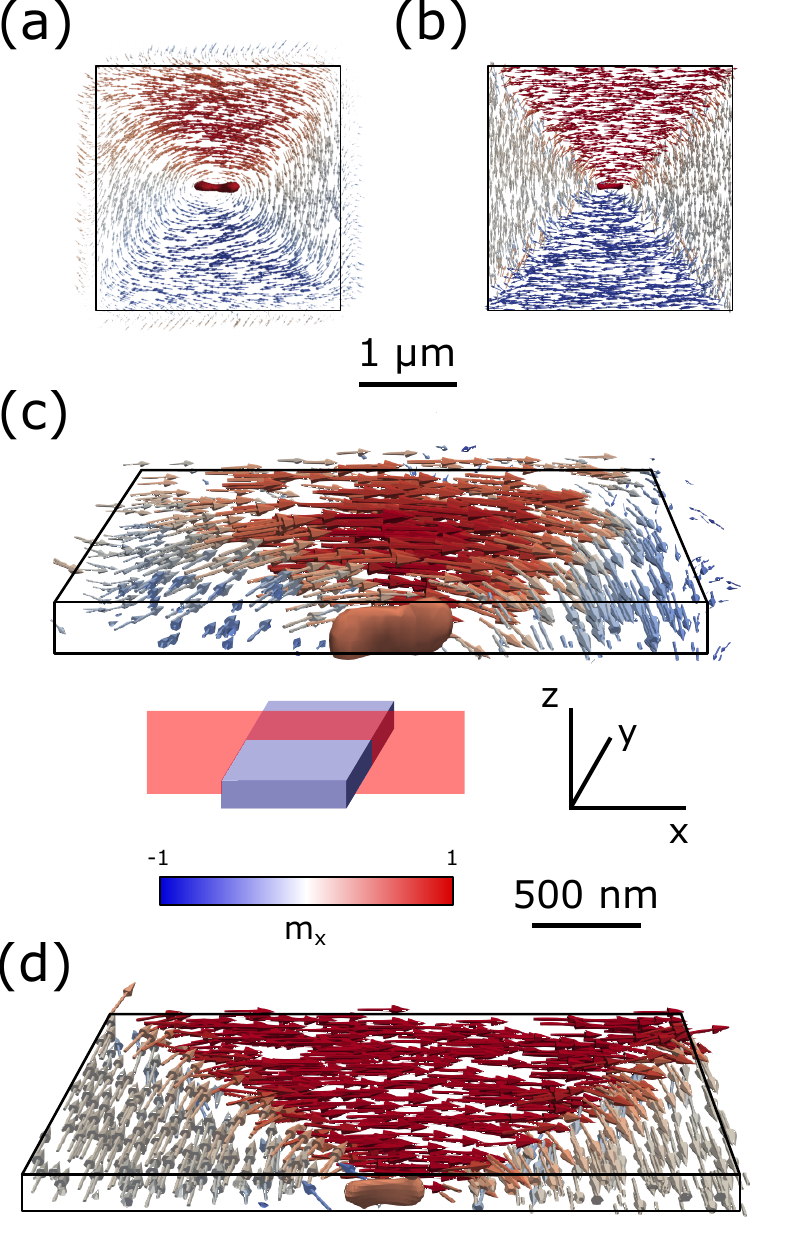}
  \caption{Three-dimensional rendering of the equilibrium configuration of the CoFeB microstructured square. (a) Top view of the reconstructed magnetic configuration, and corresponding micromagnetic simulation (b), where the red arrows mark the orientation of the magnetization in the domains. (c) Section of the measured laminogram along the the sectioning plane depicted below the image, where the orientation of the reconstructed local magnetization vector and the vortex core are shown; (d) Same section shown for a micromagnetic simulation of the CoFeB square, where the correspondence between the experimental and simulated data can be observed.}
 \label{fig:VC_quasi_static}
\end{figure}

To demonstrate the possibility of performing three-dimensional time-resolved imaging with freely selectable excitation frequencies, we performed time-resolved laminographic imaging of two spin dynamic modes of a 150 nm thick Co$_{40}$Fe$_{40}$B$_{20}$ (from now on referred to as CoFeB) 2.5$\times$2.5 $\mu$m$^2$ microstructured square, which stabilizes a flux-closure magnetic Landau state at equilibrium. Before performing the time-resolved imaging, we acquired a static magnetic laminogram of the equilibrium configuration of the CoFeB microstructure. The reconstruction of the magnitude and orientation of the local magnetization vectors was performed using the algorithm described in Ref. \cite{art:donnelly_3DReconstruction}. The reconstructed magnetic laminogram is shown in Fig. \ref{fig:VC_quasi_static} (a) and (c). Here, the arrows depict the normalized magnitude and orientation of the local magnetization vector as reconstructed from the magnetic laminogram, and are colored in a blue-white-red scale according to the $x$ component of the magnetization vector. In addition to the reconstructed local magnetic vectors, we also show the vortex core stabilized in the CoFeB microstructure, depicted as an isosurface of the $z$ component ($c_z$) of the curl of the magnetization vector $\mathbf{c} = \mathbf{\nabla} \times \mathbf{m}$ (see the Methods section for additional details).

From the static magnetic laminogram, we can observe that the CoFeB microstructured square exhibits a magnetic uniaxial anisotropy along the $x$ axis, which is evidenced by the larger area of the magnetic domains along the $x$ axis and by the shape of the vortex core. In particular, the vortex core exhibits a S-shaped structure (in contrast to the one-dimensional columnar structure that would be expected in absence of anisotropy) elongated along the anisotropy axis, with the vortex core meeting the top and bottom surfaces of the CoFeB microstructured square at two spots separated by a lateral distance of about 300 nm.

To verify that the reconstructed laminogram resembles the expected magnetic configuration, three-dimensional micromagnetic simulations of the CoFeB microstructured square with anisotropy, the value of which was determined from magneto-optical Kerr effect measurements, were performed with the finite-differences simulation package MuMax$^3$ \cite{art:mumax} (details about the simulations can be found in the Methods section). The simulated three-dimensional magnetic configuration is shown in Figs. \ref{fig:VC_quasi_static}(b) and (d), where its close resemblance to the experimental data can be observed.

%% Dynamics

Having determined the static magnetic configuration of the CoFeB microstructure, we next investigate its dynamics by performing time-resolved magnetic laminography. Specifically, by injecting a microwave signal across a Cu microantenna fabricated on top of the CoFeB microstructure, a set of different magneto-dynamical modes can be excited, ranging from the gyration of the vortex core to the excitation of the magnetic domain walls, and to the emission and propagation of spin waves. For this work, we performed three-dimensional time-resolved imaging of the fundamental vortex gyration mode (frequency of 326 MHz) and of a magnetic field-induced domain wall excitation mode (frequency of 913 MHz). The experimental setup employed for the time-resolved measurements is sketched in Fig. \ref{fig:lamni_setup}(b) and described in more detail in the Methods section.

\begin{figure}[p]
 \includegraphics{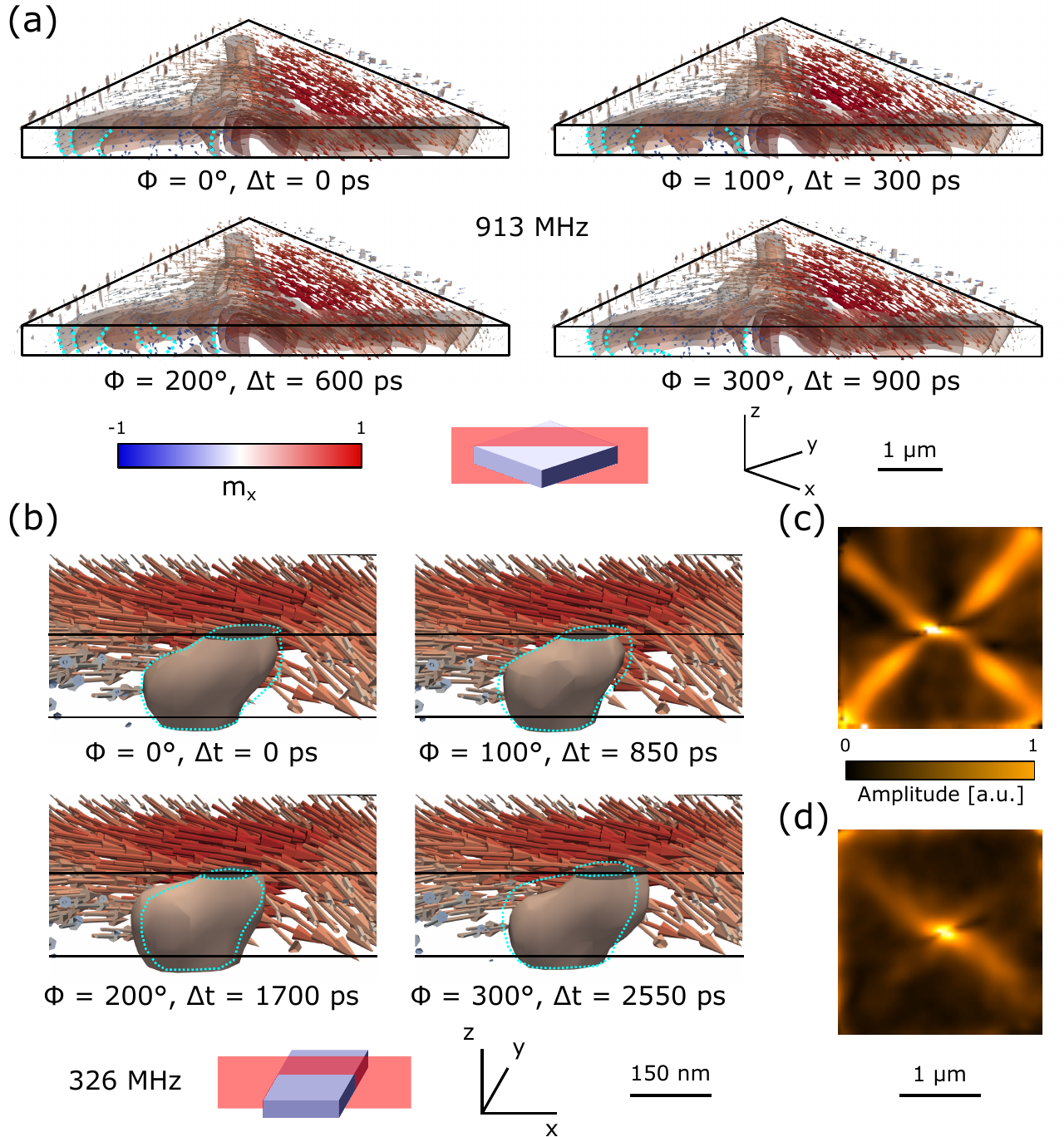}
  \caption{Three-dimensional rendering of single frames of the time-resolved magnetic laminograms acquired at an excitation frequency of (a) 913 MHz and (b) 326 MHz. Both snapshot sets show a section of the reconstructed laminogram along the red plane sketched below each snapshot set, zoomed in on the vortex core for (b). $\Phi$ indicates the phase difference between the excitation signal and the time instant probed by each snapshot. For the snapshots pictured in (a), the red isosurfaces of $c_z$ follow the domain walls. The dashed cyan lines are a guide for the eye, showing the configuration of the $c_z$ isosurfaces cut by the sectioning plane for the previous snapshot. For the snapshots pictured in (b), the vortex core is shown by the red isosurface of $c_z$. The dashed cyan lines are a guide for the eye, showing the vortex core position at the previous snapshot. (c) and (d) show the areas of the CoFeB square where the two modes are localized. (c) shows that the 913 MHz mode is localized primarily in the domain walls, while (d) shows that the 326 MHz mode is localized in the vortex core.}
 \label{fig:lamni_snapshots}
\end{figure}

Figure \ref{fig:lamni_snapshots} shows snapshots of the reconstructed magnetization profile for the domain wall excitation and vortex core gyration modes. As for the static laminogram presented in Fig. \ref{fig:VC_quasi_static}, we again show the reconstructed magnitude and orientation of the local magnetic moments and the isosurfaces of the $z$ component of the curl of the magnetization, $c_z$. For the domain wall excitation mode (Fig. \ref{fig:lamni_snapshots}(a)), we show a set of isosurfaces of $c_z$, which allow us to delineate the magnetic domain walls of the CoFeB microstructure while, for the vortex gyration mode, we show the isosurface delineating the vortex core. In addition, by determining the magnitude of the change of the magnetization across one cycle of excitation for each voxel of the time-resolved laminogram, we show the localization of the two modes. This is depicted in Fig. \ref{fig:lamni_snapshots}(c) for the 913 MHz mode, where it can be observed that the mode is localized in the domain walls of the CoFeB square, allowing it to be identified as a domain wall excitation. For the 326 MHz mode, shown in Fig. \ref{fig:lamni_snapshots}(d), instead, the mode is primarily localized in the vortex core, allowing it to be identified as a vortex gyration mode.

One of the main advantages of time-resolved laminographic imaging is that it allows us to resolve differences in the dynamics through the depth of a structure, and this can be well observed comparing the two magnetodynamical modes shown in Fig. \ref{fig:lamni_snapshots}. In particular, the vortex core gyration mode exhibits a strong variation of its dynamics across its thickness, while the domain wall excitation mode exhibits a practically uniform motion along the thickness of the CoFeB microstructure. For the domain wall excitation mode (Fig. \ref{fig:lamni_snapshots}(a)), a precession of the magnetization within the domain wall can be observed. This precession, visible by the changes in the position of the $c_z$ isosurfaces across one cycle of excitation, is predominantly uniform along the thickness of the microstructure. Therefore, while the possibility to measure this domain wall excitation mode provides a demonstration of the free selection of the excitation frequency, for the remainder of this work we will concentrate on the vortex gyration mode (Fig. \ref{fig:lamni_snapshots}(b)), where the laminographic imaging allows for the unraveling of the vortex core dynamics.

\begin{figure}[p]
 \includegraphics{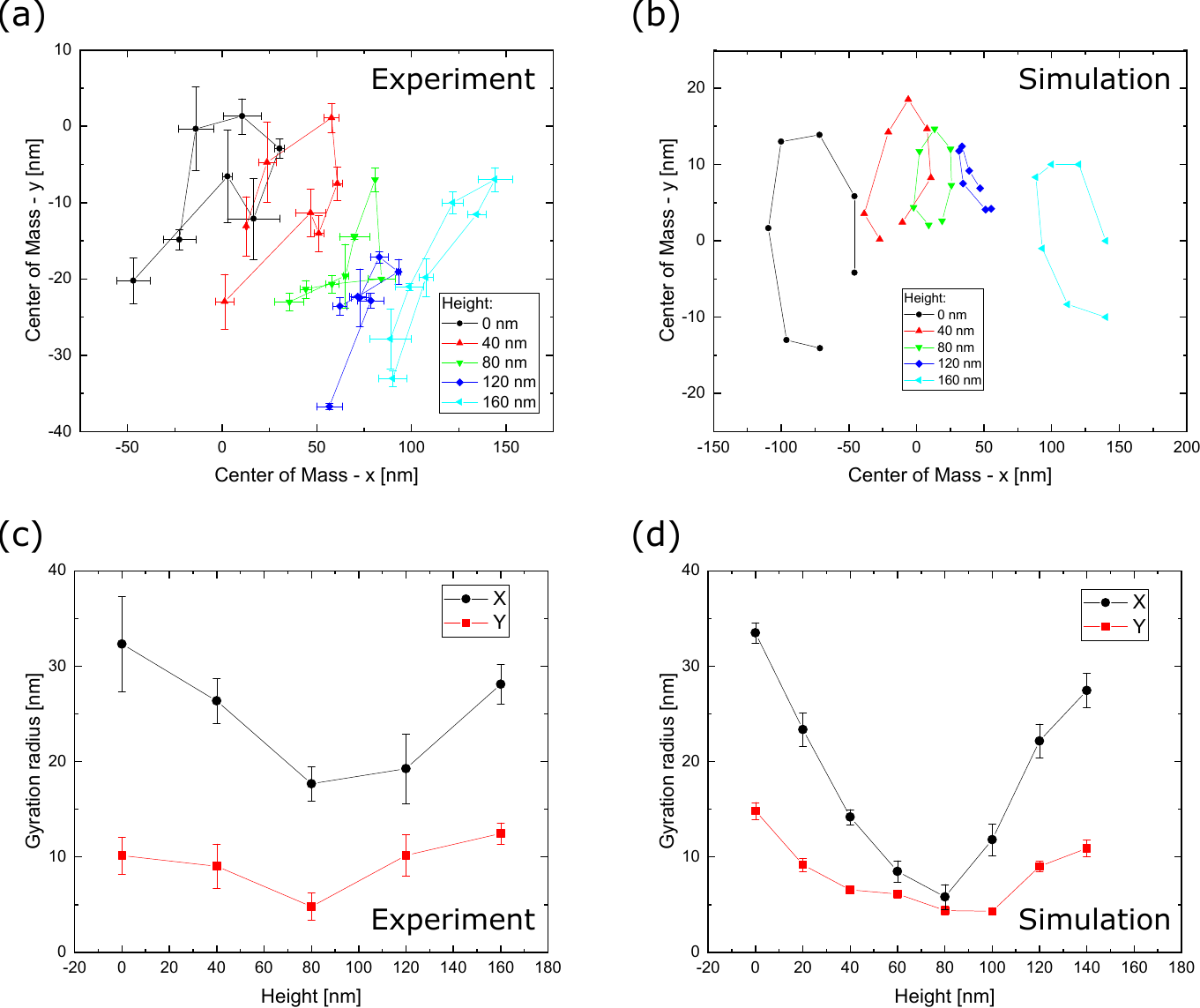}
  \caption{Experimental (a) and simulated (b) $x-y$ position of the vortex core's center of mass along the thickness of the CoFeB microstructured square, determined by calculating the center of mass of the region where $c_z \geq 0.9 \times \mathrm{max}(c_z)$ for each instant and $z$ slice of the reconstructed time-resolved laminogram.	Error bars from a combination of counting statistics and image resolution. The fitted amplitude of the $x-y$ vortex core gyration along the thickness of the CoFeB microstructured square is shown in (c) for the experimental data, and in (d) for the corresponding micromagnetic simulations. The error bars are statistical errors from the fits.}
 \label{fig:VC_gyration}
\end{figure}

The injection of the 326 MHz RF signal, close to the eigenfrequency of the fundamental vortex core gyration mode, causes the vortex core to start gyrating around its equilibrium position, shown in Fig. \ref{fig:VC_quasi_static}(a). This leads to a strong variation of its dynamical behavior across the thickness of the CoFeB microstructure, as the equilibrium configuration itself exhibits a non trivial three-dimensional structure. The motion of the vortex core can be observed in the plot shown in Fig. \ref{fig:VC_gyration}(a), where the position of the center of mass of the vortex core across each slice of the 3D time-resolved image is shown.

To obtain more insight into the three-dimensional dynamics of the vortex core, we fitted the $x$ and $y$ coordinates of the center of mass with a sinusoidal function, which allowed us to determine the amplitude of the core's gyration in the $x-y$ plane at each slice of the three-dimensional time-resolved image. This result is shown in Fig. \ref{fig:VC_gyration}(b). In particular, it can be observed that the highest amplitude in the $x-y$ core's gyration occurs at the top and bottom surfaces of the CoFeB square, where the core is perpendicular to the $x-y$ plane. This experimental result can be well-reproduced by micromagnetic simulations, as shown in Figs. \ref{fig:VC_gyration}(c) and (d).

\begin{figure}[p]
 \includegraphics{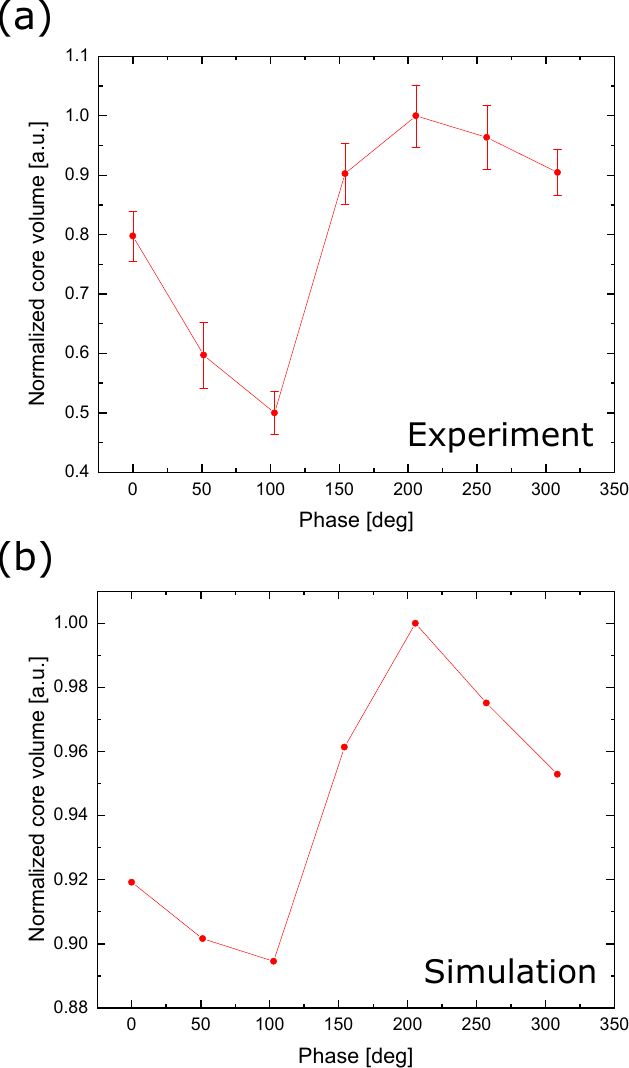}
  \caption{Temporal dependence of the vortex core volume, normalized to the maximum volume across an excitation cycle. (a) Experimental data, determined from the volume enclosed by the $0.9 \times \mathrm{max}(c_z)$ isosurface. Error bars from a combination of counting statistics and image resolution. (b) Corresponding micromagnetic simulation, showing a qualitative agreement with the experimental data.}
 \label{fig:VC_volume}
\end{figure}

The gyrotropic motion of the vortex core is also paired with a breathing mode of the core, i.e. an expansion and contraction of the vortex core volume across one cycle of excitation. The breathing of the vortex core can be visualized by determining the volume enclosed by a given isosurface of $c_z$ for each time step. The specific isosurface has been selected to be equal to 90\% of the maximum of $c_z$, and allows for the determination of the relative change in the vortex core volume across one cycle of excitation. Such time dependence is shown in Fig. \ref{fig:VC_volume}(a), and the corresponding micromagnetic simulations qualitatively confirming the experimental data are shown in Fig. \ref{fig:VC_volume}(b).

The coexistence of gyration and breathing, coupled with the three-dimensional structure of the vortex core at its equilibrium configuration provides an additional energy reservoir if compared to two dimensional vortex cores. For such three-dimensional vortex cores, this additional energy reservoir requires a generalization of the Thiele model, where a magnetic inertial mass term has to be included \cite{art:guslienko_vortexMass, art:makhfudz_gyration, art:buettner_gyration}. Through time-resolved laminographic imaging, the three-dimensional dynamical deformation of topological objects such as magnetic vortex cores can therefore be directly imaged, providing a direct experimental verification of predictions e.g. from micromagnetic simulations. This imaging protocol will also allow to reduce the reliance on micromagnetics for insights into three-dimensional dynamics, and offer the possibility to directly observe dynamical behaviors not predicted by micromagnetic simulations.

%-- Conclusions

%\section{Conclusions}

In conclusion, we have presented a new time-resolved setup based on X-ray laminography combined with time-resolved STXM imaging that allows for the investigation of magneto-dynamical processes resolved in all three spatial dimensions combined with the possibility to freely select the frequency of the excitation signal. This will allow for the possibility to resolve a range of resonant magnetization dynamic processes in three-dimensions, such as e.g. the generation, propagation, and localization of spin waves, domain wall motion dynamics, dynamics of three-dimensional topological objects such as hopfions, and the investigation of three-dimensional magnetic nanostructures. In particular, we have demonstrated here the three-dimensional mapping of two different dynamic modes that can be excited in a thick microstructured CoFeB square, stabilizing a Landau flux closure state at its equilibrium configuration, when applying oscillating magnetic fields of different frequencies. The two modes are the gyration of the vortex core, using an excitation frequency of 326 MHz, and the motion of the domain walls of the Landau pattern, obtained at a frequency of 913 MHz. The two investigated modes show a substantially different behavior along the thickness of the CoFeB square, with the domain wall excitation mode displacing uniformly along the thickness, and the vortex core gyration mode exhibiting a strongly non-uniform behavior along its thickness, which is further enhanced by the uniaxial anisotropy of the CoFeB microstructure. For this mode, the counterclockwise gyration of the vortex core around its equilibrium position is combined with a breathing of the vortex core, indicating that energy can be stored in the deformation of the core itself, following the behavior observed in other topological structures such as the magnetic skyrmion bubble. 

As a final note, it should be considered that, for time-resolved laminography imaging, a set of projections has to be acquired in order to reconstruct the three-dimensional magnetization configuration. Due to the requirement of acquiring three polarizations (circular positive, negative, and linear horizontal) to reconstruct the magnetic configuration \cite{art:witte_Lamni}, to the requirement of more than 30-40 projections to obtain a sufficiently good $z$ resolution \cite{art:holler_lamni}, and to the fact that a time-resolved image is composed of N frames (typically on the order of 10 for RF dynamics), all of which requiring sufficient statistics \cite{art:finizio_TR_STXM_simulations}, a time-resolved laminography image currently requires a considerable investment in measurement time. Nonetheless, thanks to the planned upgrades of several synchrotrons towards diffraction limited sources, the future of time-resolved laminography imaging is bright: improvements in the intensity, brilliance, and coherence of the X-ray beam produced by diffraction limited source will allow for the reduction of integration times of several factors.%% The new capabilities offered by time-resolved soft X-ray laminographic imaging, combined with brighter X-ray sources, will open up the possibility to investigate a range of resonant magneto-dynamical processes in three dimensions, including e.g. the generation and propagation of spin waves, domain wall motion processes, dynamics of three-dimensional structures such as magnetic hopfions and vortex rings \cite{art:donnelly_vortexRing}, and of three-dimensional magnetic nanostructures.

\section{Methods}

\subsection{Sample fabrication}

Microstructured CoFeB squares (2.5 $\mathrm{\mu}$m wide, 150 nm thick) were patterned on top of a 200 nm thick, 1$\times$1 mm$^2$ wide Si$_3$N$_4$ membrane on a 5$\times$5 mm$^2$ intrinsic Si frame (250 $\mathrm{\mu}$m thick, from Silson Ltd.). The patterning was performed by electron beam lithography using a Vistec EBPG5000, 100 kV electron beam writer. A bilayer resist composed of a layer of poly(methyl-methacrylate) (4\% in Anisole, 3000 rpm for 1 minute) on top of a layer of methyl methacrylate (6\% in Ethyl-Lactate, 3000 rpm for 1 minute) was spincoated on top of the Si$_3$N$_4$ membrane, followed by a 1 minute soft-bake at 175 $^\circ$C for both resist layers. Following the spincoating, the resist was exposed with a dose of 1500 $\mathrm{\mu}$C cm$^{-2}$. Following development by immersion in a 1:3 (volume) solution of methyl isobutyl ketone and isopropyl alcohol (IPA) and rinsing in pure IPA (each for 45 seconds), a 150 nm thick CoFeB film was deposited by DC sputtering using an AJA sputter coater, where a 2 nm thick Ta seed layer was deposited prior to the CoFeB. The unexposed resist, together with the CoFeB film on top of it was then removed by immersion of the sample in pure acetone, leaving only the patterned CoFeB square.

On top of the CoFeB square, a 5 $\mu$m wide, 300 nm thick, Cu stripline designed to allow for the generation of an oscillating in-plane magnetic field along the $x$ axis was fabricated, again by electron beam lithography followed by liftoff using the same recipe as for the CoFeB square. The Cu film was deposited by thermal evaporation from a pure Cu pellet, using a Balzers BAE evaporator. The quality of the fabricated samples was verified by scanning electron microscopy and time-domain reflectometry measurements.

\subsection{Time-resolved soft X-ray laminography}

The three-dimensional images were acquired following the laminography imaging protocol \cite{art:holler_lamni}, where a set of two-dimensional projections at different rotation angles of the sample was acquired. The single projections were acquired by STXM imaging, utilizing diffractive optics (Fresnel zone plate) to focus the X-ray beam. Due to geometrical constraints, a laminography angle of 45$^\circ$ was employed, combined with a Fresnel zone plate with a 50 nm outermost zone width, which defines the achievable spatial resolution. The X-ray imaging was performed at the L$_2$ absorption edge of Cobalt, corresponding to an X-ray energy of about 793 eV. The L$_2$ edge was selected rather than the more common L$_3$ edge due to the lower X-ray absorption at the L$_2$ edge, which is a critical factor when considering the effective CoFeB thickness of about 210 nm under a laminography angle of 45$^\circ$.

A total of 50 projections were acquired for each time-resolved laminography image, resulting in a theoretical maximum $z$ resolution of 10 nm, calculated according to the following equation \cite{art:holler_lamni}:
\begin{equation*}
P \geq \frac{\pi d}{\Delta r} \tan(\theta),
\end{equation*}
where $\Delta r$ identifies the z resolution, $d$ the thickness of the CoFeB structure (150 nm), $P$ the number of projections (50), and $\theta$ the laminography angle (45$^\circ$).

For each projection, three time-resolved STXM images were acquired, two with circularly polarized X-rays of opposite helicities, and the third with linearly polarized X-rays (linear horizontal). This is a different approach to standard time-resolved two-dimensional STXM imaging, where only a single image with circularly polarized X-rays is acquired (e.g. \cite{art:wintz_spin_waves}), and is necessary to allow for the reconstruction of the three-dimensional orientation of the local magnetization vectors \cite{art:donnelly_TRLamni}. The acquisition of an image with linearly polarized X-rays is required to allow for the correction of possible discrepancies in the degree of circular polarization that can occur when performing XMCD imaging at a bending magnet beamline such as PolLux \cite{art:pollux, art:witte_Lamni}.

For the time-resolved imaging, a set of 23 frames for each projection was acquired. To improve the statistics of the single frames, the 23 raw frames were binned (average) to 7 frames. The dynamics were excited by the injection of a RF current across the Cu stripline fabricated on top of the CoFeB microstructured square, giving rise to an oscillating magnetic field through the Oersted effect. The amplitude of the RF current was selected in order to generate an oscillating magnetic field with a 5 mT amplitude. The RF current was generated by a Keysight M8195A arbitrary waveform generator (64 GSa s$^{-1}$ bandwidth) frequency locked to the 500 MHz master clock of the synchrotron light source, combined with a SHF826H broadband amplifier. The electrical circuit is schematically depicted in Fig. \ref{fig:lamni_setup}(b). The frequency of the excitation signal was selected according to the relation $f = 500 \frac{M}{N}$ [MHz], being $N$ equal to the number of raw frames in the image (23) and $M$ an integer multiplier \cite{art:puzic_TR_STXM}. The frequency closest to the vortex gyration eigenfrequency was attained for $M = 15$ (326 MHz), and the domain wall excitation mode was triggered at $M = 42$ (913 MHz). These frequencies were found by two-dimensional TR-STXM imaging, where a set of time-resolved images as a function of the applied excitation frequency were acquired. The temporal resolution of the measurements presented here is determined by the width of the X-ray pulses generated by the synchrotron, which is of 70 ps FWHM for the Swiss Light Source operating with a multibunch filling pattern \cite{art:finizio_QuTAG}.

To contact the stripline, a custom-designed printed circuit board (PCB), designed to be compatible with the constraints of the laminography stage while still guaranteeing good RF performances, was utilized. A Boa-Flex III flexible RF cable from Huber+Suhner was utilized for the electrical connection of the PCB. The high flexibility of the cable enabled for the possibility to perform several rotations of the laminography stage whilst maintaining an electrical connection to the PCB, and good RF properties.

The reconstruction of the three-dimensional topographic and magnetic configuration of the CoFeB square was then performed for each of the 7 binned frames according to the process described in \cite{art:donnelly_TRLamni, art:witte_Lamni}. The final three-dimensional time-resolved images were then rendered using the free software Paraview \cite{book:paraview}. There, the orientation and magnitude of the magnetic moment in each voxel of the three-dimensional image were rendered, using a mask defined by the reconstructed topography of the CoFeB square. To visualize the vortex core, we also plot the magnitude of the $z$ component, $c_z$, of the curl of the magnetization (calculated as $\mathbf{c} = \mathbf{\nabla} \times \mathbf{m}$, being $\mathbf{m}$ the local magnetization vector). The $c_z$ component provides an easy method for the direct visualization of the vortex core, as shown in Ref. \cite{art:donnelly_TRLamni} and as demonstrated from comparison with micromagnetic simulations (see Fig. \ref{fig:VC_quasi_static} and the supplementary information).

The localization of the two modes investigated here, shown in Fig. \ref{fig:lamni_snapshots}(c-d), was determined by calculating, for each voxel in the time-resolved laminogram, a fast Fourier transform, and plotting the magnitude of the Fourier transform at the frequency of the mode (913 MHz for Fig. \ref{fig:lamni_snapshots}(c) and 326 MHz for Fig. \ref{fig:lamni_snapshots}(d)).

The vortex core position along the thickness of the microstructured CoFeB square was determined by calculating the center of mass of the regions where $c_z \geq 0.9 \times \mathrm{max}(c_z)$ for each of the $z$ slices of the reconstructed laminogram. The calculation of the center of mass was not weighted by the value of $c_z$ (i.e. all pixels where $c_z \geq 0.9 \times \mathrm{max}(c_z)$ have the same weight in the calculation). For the micromagnetic simulations, the center of mass of the vortex core was determined by calculating the center of mass of the region where $m_z > 0.9$, being $m_z$ the normalized $z$ component of the magnetization. To allow for an easier comparison between the simulations and the experimental data, the simulated data was subdivided in 8 slices.

The volume of the vortex core was determined by calculating the volume surrounded by the $c_z = 0.9 \times{max}(c_z)$ isosurface. For the micromagnetic simulations, this calculation was performed by considering the volume enclosed by the $m_z = 0.9$ isosurface.

\subsection{Micromagnetic simulations}

Micromagnetic simulations of the static and dynamical behavior of the CoFeB microstructured square presented here were performed by numerically solving the Landau-Lifshitz-Gilbert equation using the finite differences MuMax$^3$ framework \cite{art:mumax}. A three-dimensional simulation grid was considered, consisting of a 512 $\times$ 512 $\times$ 32 px$^3$ lattice with 4.88 $\times$ 4.88 $\times$ 4.68 nm$^3$ cell. We selected a smaller step size along the $z$ direction to keep a power of 2 as the number of cells for all directions, allowing for a more efficient computation. As micromagnetic parameters, we employed a saturation magnetization of M$_\mathrm{s}$ = 10$^6$ A m$^{-1}$, an exchange stiffness of A$_\mathrm{ex}$ = 10$^{-11}$ J m$^{-1}$, a Gilbert damping constant of $\alpha$ = 0.05. These micromagnetic parameters result in an exchange length of about 4 nm. To verify that the size of the discretization is not affecting the results of the micromagnetic simulations, a subset of the simulations was performed with a 2.44 $\times$ 2.44 $\times$ 2.34 nm$^3$ discretization, and the results of the simulation with the finer grid closely resemble the simulations performed with the larger grid.

For all of the time-resolved simulations, the starting configuration was determined by relaxing a symmetric Landau pattern under an applied uniaxial anisotropy of 5 kJ m$^{-3}$, the value of which was determined by magneto-optical Kerr effect measurements, and verified by comparing a static micromagnetic simulation with the observed magnetic configuration. A sinusoidal magnetic field along the direction of the uniaxial anisotropy ($x$ axis) was then applied, reproducing the experimental configuration. The amplitude of the magnetic field was 5 mT, and a 200 ns period was simulated prior to recording the simulated data to allow for transient effects caused by the sudden application of the magnetic field to dissipate.

\section{Author Contributions}
SF conceived the experiment with input from CD and JR. SF and JR implemented the time-resolved laminography setup. SF designed the sample and performed the lithographical patterning. AH deposited the CoFeB films. SF, SM, CD, and AH performed the time-resolved laminography experiments and interpreted the resulting data, using code provided by CD. SF performed the micromagnetic simulations. SF wrote the manuscript, with input from all authors.

\section{Acknowledgments}

This work was performed at the PolLux (X07DA) endstation of the Swiss Light Source, Paul Scherrer Institut, Villigen PSI, Switzerland. The research leading to these results has received funding from the Swiss National Science Foundation under grant agreement No. 172517. The PolLux endstation was financed by the German Bundesministerium f\"ur Bildung und Forschung through contracts 05K16WED and 05K19WE2. C.D. acknowledges support from the Max Planck Society Lise Meitner Excellence Program.

\section{Supporting Information}

The following supporting information is available:
\begin{itemize}
 \item[-] \textbf{DomainWallExcitation.avi} - Video displaying the change in the spin configuration and the motion of the magnetic domain walls for the domain wall excitation mode (913 MHz). The video is shown in the same perspective as for Fig. \ref{fig:lamni_snapshots}(a).
 \item[-] \textbf{VortexGyration.avi} - Video displaying the change in the spin configuration and the deformation of the vortex core for the vortex core gyration mode (326 MHz). The video is shown in the same perspective as for Fig. \ref{fig:lamni_snapshots}(b).
\end{itemize}


\begin{thebibliography}{30}%
\makeatletter
\providecommand \@ifxundefined [1]{%
 \@ifx{#1\undefined}
}%
\providecommand \@ifnum [1]{%
 \ifnum #1\expandafter \@firstoftwo
 \else \expandafter \@secondoftwo
 \fi
}%
\providecommand \@ifx [1]{%
 \ifx #1\expandafter \@firstoftwo
 \else \expandafter \@secondoftwo
 \fi
}%
\providecommand \natexlab [1]{#1}%
\providecommand \enquote  [1]{``#1''}%
\providecommand \bibnamefont  [1]{#1}%
\providecommand \bibfnamefont [1]{#1}%
\providecommand \citenamefont [1]{#1}%
\providecommand \href@noop [0]{\@secondoftwo}%
\providecommand \href [0]{\begingroup \@sanitize@url \@href}%
\providecommand \@href[1]{\@@startlink{#1}\@@href}%
\providecommand \@@href[1]{\endgroup#1\@@endlink}%
\providecommand \@sanitize@url [0]{\catcode `\\12\catcode `\$12\catcode
  `\&12\catcode `\#12\catcode `\^12\catcode `\_12\catcode `\%12\relax}%
\providecommand \@@startlink[1]{}%
\providecommand \@@endlink[0]{}%
\providecommand \url  [0]{\begingroup\@sanitize@url \@url }%
\providecommand \@url [1]{\endgroup\@href {#1}{\urlprefix }}%
\providecommand \urlprefix  [0]{URL }%
\providecommand \Eprint [0]{\href }%
\providecommand \doibase [0]{http://dx.doi.org/}%
\providecommand \selectlanguage [0]{\@gobble}%
\providecommand \bibinfo  [0]{\@secondoftwo}%
\providecommand \bibfield  [0]{\@secondoftwo}%
\providecommand \translation [1]{[#1]}%
\providecommand \BibitemOpen [0]{}%
\providecommand \bibitemStop [0]{}%
\providecommand \bibitemNoStop [0]{.\EOS\space}%
\providecommand \EOS [0]{\spacefactor3000\relax}%
\providecommand \BibitemShut  [1]{\csname bibitem#1\endcsname}%
\let\auto@bib@innerbib\@empty
%</preamble>
\bibitem [{\citenamefont {Guslienko}(2006)}]{art:guslienko_gyrovector}%
  \BibitemOpen
  \bibfield  {author} {\bibinfo {author} {\bibfnamefont {K.~Y.}\ \bibnamefont
  {Guslienko}},\ }\href@noop {} {\bibfield  {journal} {\bibinfo  {journal}
  {Applied Physics Letters}\ }\textbf {\bibinfo {volume} {89}},\ \bibinfo
  {pages} {022510} (\bibinfo {year} {2006})}\BibitemShut {NoStop}%
\bibitem [{\citenamefont {Guslienko}\ \emph {et~al.}(2015)\citenamefont
  {Guslienko}, \citenamefont {Kakazei}, \citenamefont {Ding}, \citenamefont
  {Liu},\ and\ \citenamefont {Adeyeye}}]{art:guslienko_vortexMass}%
  \BibitemOpen
  \bibfield  {author} {\bibinfo {author} {\bibfnamefont {K.~Y.}\ \bibnamefont
  {Guslienko}}, \bibinfo {author} {\bibfnamefont {G.~N.}\ \bibnamefont
  {Kakazei}}, \bibinfo {author} {\bibfnamefont {J.}~\bibnamefont {Ding}},
  \bibinfo {author} {\bibfnamefont {X.~M.}\ \bibnamefont {Liu}}, \ and\
  \bibinfo {author} {\bibfnamefont {A.~O.}\ \bibnamefont {Adeyeye}},\
  }\href@noop {} {\bibfield  {journal} {\bibinfo  {journal} {Scientific
  Reports}\ }\textbf {\bibinfo {volume} {5}},\ \bibinfo {pages} {13881}
  (\bibinfo {year} {2015})}\BibitemShut {NoStop}%
\bibitem [{\citenamefont {{Kr\"uger}}\ \emph {et~al.}(2007)\citenamefont
  {{Kr\"uger}}, \citenamefont {Drews}, \citenamefont {Bolte}, \citenamefont
  {Merkt}, \citenamefont {Pfannkuche},\ and\ \citenamefont
  {Meier}}]{art:krueger_harmonicPotential}%
  \BibitemOpen
  \bibfield  {author} {\bibinfo {author} {\bibfnamefont {B.}~\bibnamefont
  {{Kr\"uger}}}, \bibinfo {author} {\bibfnamefont {A.}~\bibnamefont {Drews}},
  \bibinfo {author} {\bibfnamefont {M.}~\bibnamefont {Bolte}}, \bibinfo
  {author} {\bibfnamefont {U.}~\bibnamefont {Merkt}}, \bibinfo {author}
  {\bibfnamefont {D.}~\bibnamefont {Pfannkuche}}, \ and\ \bibinfo {author}
  {\bibfnamefont {G.}~\bibnamefont {Meier}},\ }\href@noop {} {\bibfield
  {journal} {\bibinfo  {journal} {Physical Review B}\ }\textbf {\bibinfo
  {volume} {76}},\ \bibinfo {pages} {224426} (\bibinfo {year}
  {2007})}\BibitemShut {NoStop}%
\bibitem [{\citenamefont {Dieterle}\ \emph {et~al.}(2019)\citenamefont
  {Dieterle}, \citenamefont {Foerster}, \citenamefont {Stoll}, \citenamefont
  {Semisalova}, \citenamefont {Finizio}, \citenamefont {Gangwar}, \citenamefont
  {Weigand}, \citenamefont {Noske}, \citenamefont {Faehnle}, \citenamefont
  {Bykova}, \citenamefont {Graefe}, \citenamefont {Bozhko}, \citenamefont
  {Musiienko-Shmarova}, \citenamefont {Tiberkevich}, \citenamefont {Slavin},
  \citenamefont {Back}, \citenamefont {Raabe}, \citenamefont {Schuetz},\ and\
  \citenamefont {Wintz}}]{art:dieterle_SpinWaves}%
  \BibitemOpen
  \bibfield  {author} {\bibinfo {author} {\bibfnamefont {G.}~\bibnamefont
  {Dieterle}}, \bibinfo {author} {\bibfnamefont {J.}~\bibnamefont {Foerster}},
  \bibinfo {author} {\bibfnamefont {H.}~\bibnamefont {Stoll}}, \bibinfo
  {author} {\bibfnamefont {A.~S.}\ \bibnamefont {Semisalova}}, \bibinfo
  {author} {\bibfnamefont {S.}~\bibnamefont {Finizio}}, \bibinfo {author}
  {\bibfnamefont {A.}~\bibnamefont {Gangwar}}, \bibinfo {author} {\bibfnamefont
  {M.}~\bibnamefont {Weigand}}, \bibinfo {author} {\bibfnamefont
  {M.}~\bibnamefont {Noske}}, \bibinfo {author} {\bibfnamefont
  {M.}~\bibnamefont {Faehnle}}, \bibinfo {author} {\bibfnamefont
  {I.}~\bibnamefont {Bykova}}, \bibinfo {author} {\bibfnamefont
  {J.}~\bibnamefont {Graefe}}, \bibinfo {author} {\bibfnamefont
  {A.}~\bibnamefont {Bozhko}}, \bibinfo {author} {\bibfnamefont {H.~Y.}\
  \bibnamefont {Musiienko-Shmarova}}, \bibinfo {author} {\bibfnamefont
  {V.}~\bibnamefont {Tiberkevich}}, \bibinfo {author} {\bibfnamefont {A.~N.}\
  \bibnamefont {Slavin}}, \bibinfo {author} {\bibfnamefont {C.~H.}\
  \bibnamefont {Back}}, \bibinfo {author} {\bibfnamefont {J.}~\bibnamefont
  {Raabe}}, \bibinfo {author} {\bibfnamefont {G.}~\bibnamefont {Schuetz}}, \
  and\ \bibinfo {author} {\bibfnamefont {S.}~\bibnamefont {Wintz}},\
  }\href@noop {} {\bibfield  {journal} {\bibinfo  {journal} {Physical Review
  Letters}\ }\textbf {\bibinfo {volume} {122}},\ \bibinfo {pages} {117202}
  (\bibinfo {year} {2019})}\BibitemShut {NoStop}%
\bibitem [{\citenamefont {Wintz}\ \emph {et~al.}(2016)\citenamefont {Wintz},
  \citenamefont {Tiberkevich}, \citenamefont {Weigand}, \citenamefont {Raabe},
  \citenamefont {Lindner}, \citenamefont {Erbe}, \citenamefont {Slavin},\ and\
  \citenamefont {Fassbender}}]{art:wintz_spin_waves}%
  \BibitemOpen
  \bibfield  {author} {\bibinfo {author} {\bibfnamefont {S.}~\bibnamefont
  {Wintz}}, \bibinfo {author} {\bibfnamefont {V.}~\bibnamefont {Tiberkevich}},
  \bibinfo {author} {\bibfnamefont {M.}~\bibnamefont {Weigand}}, \bibinfo
  {author} {\bibfnamefont {J.}~\bibnamefont {Raabe}}, \bibinfo {author}
  {\bibfnamefont {J.}~\bibnamefont {Lindner}}, \bibinfo {author} {\bibfnamefont
  {A.}~\bibnamefont {Erbe}}, \bibinfo {author} {\bibfnamefont {A.}~\bibnamefont
  {Slavin}}, \ and\ \bibinfo {author} {\bibfnamefont {J.}~\bibnamefont
  {Fassbender}},\ }\href@noop {} {\bibfield  {journal} {\bibinfo  {journal}
  {Nature Nanotechnology}\ }\textbf {\bibinfo {volume} {11}},\ \bibinfo {pages}
  {948} (\bibinfo {year} {2016})}\BibitemShut {NoStop}%
\bibitem [{\citenamefont {Mayr}\ \emph {et~al.}(2021)\citenamefont {Mayr},
  \citenamefont {Flajsman}, \citenamefont {Finizio}, \citenamefont {Hrabec},
  \citenamefont {Weigand}, \citenamefont {F{\"o}rster}, \citenamefont {Stoll},
  \citenamefont {Heyderman}, \citenamefont {Urbanek},\ and\ \citenamefont
  {Wintz}}]{art:mayr_spinwaves}%
  \BibitemOpen
  \bibfield  {author} {\bibinfo {author} {\bibfnamefont {S.}~\bibnamefont
  {Mayr}}, \bibinfo {author} {\bibfnamefont {L.}~\bibnamefont {Flajsman}},
  \bibinfo {author} {\bibfnamefont {S.}~\bibnamefont {Finizio}}, \bibinfo
  {author} {\bibfnamefont {A.}~\bibnamefont {Hrabec}}, \bibinfo {author}
  {\bibfnamefont {M.}~\bibnamefont {Weigand}}, \bibinfo {author} {\bibfnamefont
  {J.}~\bibnamefont {F{\"o}rster}}, \bibinfo {author} {\bibfnamefont
  {H.}~\bibnamefont {Stoll}}, \bibinfo {author} {\bibfnamefont {L.~J.}\
  \bibnamefont {Heyderman}}, \bibinfo {author} {\bibfnamefont {M.}~\bibnamefont
  {Urbanek}}, \ and\ \bibinfo {author} {\bibfnamefont {S.}~\bibnamefont
  {Wintz}},\ }\href@noop {} {\bibfield  {journal} {\bibinfo  {journal} {Nano
  Letters}\ }\textbf {\bibinfo {volume} {21}},\ \bibinfo {pages} {1584}
  (\bibinfo {year} {2021})}\BibitemShut {NoStop}%
\bibitem [{\citenamefont {Divinskiy}\ \emph {et~al.}(2021)\citenamefont
  {Divinskiy}, \citenamefont {Merbouche}, \citenamefont {Demidov},
  \citenamefont {Nikolaev}, \citenamefont {Soumah}, \citenamefont {Gorere},
  \citenamefont {Lebrun}, \citenamefont {Cros}, \citenamefont {{Ben Youssef}},
  \citenamefont {Bortolotti}, \citenamefont {Anane},\ and\ \citenamefont
  {Demokritov}}]{art:divinskiy_BEC_SOT}%
  \BibitemOpen
  \bibfield  {author} {\bibinfo {author} {\bibfnamefont {B.}~\bibnamefont
  {Divinskiy}}, \bibinfo {author} {\bibfnamefont {H.}~\bibnamefont
  {Merbouche}}, \bibinfo {author} {\bibfnamefont {V.~E.}\ \bibnamefont
  {Demidov}}, \bibinfo {author} {\bibfnamefont {K.~O.}\ \bibnamefont
  {Nikolaev}}, \bibinfo {author} {\bibfnamefont {L.}~\bibnamefont {Soumah}},
  \bibinfo {author} {\bibfnamefont {D.}~\bibnamefont {Gorere}}, \bibinfo
  {author} {\bibfnamefont {R.}~\bibnamefont {Lebrun}}, \bibinfo {author}
  {\bibfnamefont {V.}~\bibnamefont {Cros}}, \bibinfo {author} {\bibfnamefont
  {J.}~\bibnamefont {{Ben Youssef}}}, \bibinfo {author} {\bibfnamefont
  {P.}~\bibnamefont {Bortolotti}}, \bibinfo {author} {\bibfnamefont
  {A.}~\bibnamefont {Anane}}, \ and\ \bibinfo {author} {\bibfnamefont {S.~O.}\
  \bibnamefont {Demokritov}},\ }\href@noop {} {\bibfield  {journal} {\bibinfo
  {journal} {Nature Communications}\ }\textbf {\bibinfo {volume} {12}},\
  \bibinfo {pages} {6541} (\bibinfo {year} {2021})}\BibitemShut {NoStop}%
\bibitem [{\citenamefont {Jue}\ \emph {et~al.}(2016)\citenamefont {Jue},
  \citenamefont {Thiaville}, \citenamefont {Pizzini}, \citenamefont {Miltat},
  \citenamefont {Sampaio}, \citenamefont {Buda-Prejebeanu}, \citenamefont
  {Rohart}, \citenamefont {Vogel}, \citenamefont {Bonfim}, \citenamefont
  {Boulle}, \citenamefont {Auffret}, \citenamefont {Miron},\ and\ \citenamefont
  {Gaudin}}]{art:jue_FIDWM}%
  \BibitemOpen
  \bibfield  {author} {\bibinfo {author} {\bibfnamefont {E.}~\bibnamefont
  {Jue}}, \bibinfo {author} {\bibfnamefont {A.}~\bibnamefont {Thiaville}},
  \bibinfo {author} {\bibfnamefont {S.}~\bibnamefont {Pizzini}}, \bibinfo
  {author} {\bibfnamefont {J.}~\bibnamefont {Miltat}}, \bibinfo {author}
  {\bibfnamefont {J.}~\bibnamefont {Sampaio}}, \bibinfo {author} {\bibfnamefont
  {L.}~\bibnamefont {Buda-Prejebeanu}}, \bibinfo {author} {\bibfnamefont
  {S.}~\bibnamefont {Rohart}}, \bibinfo {author} {\bibfnamefont
  {J.}~\bibnamefont {Vogel}}, \bibinfo {author} {\bibfnamefont
  {M.}~\bibnamefont {Bonfim}}, \bibinfo {author} {\bibfnamefont
  {O.}~\bibnamefont {Boulle}}, \bibinfo {author} {\bibfnamefont
  {S.}~\bibnamefont {Auffret}}, \bibinfo {author} {\bibfnamefont
  {I.}~\bibnamefont {Miron}}, \ and\ \bibinfo {author} {\bibfnamefont
  {G.}~\bibnamefont {Gaudin}},\ }\href@noop {} {\bibfield  {journal} {\bibinfo
  {journal} {Physical Review B}\ }\textbf {\bibinfo {volume} {93}},\ \bibinfo
  {pages} {014403} (\bibinfo {year} {2016})}\BibitemShut {NoStop}%
\bibitem [{\citenamefont {Rhensius}\ \emph {et~al.}(2010)\citenamefont
  {Rhensius}, \citenamefont {Heyne}, \citenamefont {Backes}, \citenamefont
  {Krzyk}, \citenamefont {Heyderman}, \citenamefont {Joly}, \citenamefont
  {Nolting},\ and\ \citenamefont
  {Kl{\"a}ui}}]{art:rhensius_field_domain_motion}%
  \BibitemOpen
  \bibfield  {author} {\bibinfo {author} {\bibfnamefont {J.}~\bibnamefont
  {Rhensius}}, \bibinfo {author} {\bibfnamefont {L.}~\bibnamefont {Heyne}},
  \bibinfo {author} {\bibfnamefont {D.}~\bibnamefont {Backes}}, \bibinfo
  {author} {\bibfnamefont {S.}~\bibnamefont {Krzyk}}, \bibinfo {author}
  {\bibfnamefont {L.~J.}\ \bibnamefont {Heyderman}}, \bibinfo {author}
  {\bibfnamefont {L.}~\bibnamefont {Joly}}, \bibinfo {author} {\bibfnamefont
  {F.}~\bibnamefont {Nolting}}, \ and\ \bibinfo {author} {\bibfnamefont
  {M.}~\bibnamefont {Kl{\"a}ui}},\ }\href@noop {} {\bibfield  {journal}
  {\bibinfo  {journal} {Physical Review Letters}\ }\textbf {\bibinfo {volume}
  {104}},\ \bibinfo {pages} {067201} (\bibinfo {year} {2010})}\BibitemShut
  {NoStop}%
\bibitem [{\citenamefont {Ryu}\ \emph {et~al.}(2013)\citenamefont {Ryu},
  \citenamefont {Thomas}, \citenamefont {Yang},\ and\ \citenamefont
  {Parkin}}]{art:ryu_CIDWM}%
  \BibitemOpen
  \bibfield  {author} {\bibinfo {author} {\bibfnamefont {K.-S.}\ \bibnamefont
  {Ryu}}, \bibinfo {author} {\bibfnamefont {L.}~\bibnamefont {Thomas}},
  \bibinfo {author} {\bibfnamefont {S.-H.}\ \bibnamefont {Yang}}, \ and\
  \bibinfo {author} {\bibfnamefont {S.}~\bibnamefont {Parkin}},\ }\href@noop {}
  {\bibfield  {journal} {\bibinfo  {journal} {Nature Nanotechnology}\ }\textbf
  {\bibinfo {volume} {8}},\ \bibinfo {pages} {527} (\bibinfo {year}
  {2013})}\BibitemShut {NoStop}%
\bibitem [{\citenamefont {Litzius}\ \emph {et~al.}(2017)\citenamefont
  {Litzius}, \citenamefont {Lemesh}, \citenamefont {Kr{\"u}ger}, \citenamefont
  {Bassirian}, \citenamefont {Caretta}, \citenamefont {Richter}, \citenamefont
  {B{\"u}ttner}, \citenamefont {Sato}, \citenamefont {Tretiakov}, \citenamefont
  {F{\"o}rster}, \citenamefont {Reeve}, \citenamefont {Weigand}, \citenamefont
  {Bykova}, \citenamefont {Stoll}, \citenamefont {Sch{\"u}tz}, \citenamefont
  {Beach},\ and\ \citenamefont {Kl{\"{a}}ui}}]{art:kai_skyrmion_hall_angle}%
  \BibitemOpen
  \bibfield  {author} {\bibinfo {author} {\bibfnamefont {K.}~\bibnamefont
  {Litzius}}, \bibinfo {author} {\bibfnamefont {I.}~\bibnamefont {Lemesh}},
  \bibinfo {author} {\bibfnamefont {B.}~\bibnamefont {Kr{\"u}ger}}, \bibinfo
  {author} {\bibfnamefont {P.}~\bibnamefont {Bassirian}}, \bibinfo {author}
  {\bibfnamefont {L.}~\bibnamefont {Caretta}}, \bibinfo {author} {\bibfnamefont
  {K.}~\bibnamefont {Richter}}, \bibinfo {author} {\bibfnamefont
  {F.}~\bibnamefont {B{\"u}ttner}}, \bibinfo {author} {\bibfnamefont
  {K.}~\bibnamefont {Sato}}, \bibinfo {author} {\bibfnamefont {O.~A.}\
  \bibnamefont {Tretiakov}}, \bibinfo {author} {\bibfnamefont {J.}~\bibnamefont
  {F{\"o}rster}}, \bibinfo {author} {\bibfnamefont {R.~M.}\ \bibnamefont
  {Reeve}}, \bibinfo {author} {\bibfnamefont {M.}~\bibnamefont {Weigand}},
  \bibinfo {author} {\bibfnamefont {I.}~\bibnamefont {Bykova}}, \bibinfo
  {author} {\bibfnamefont {H.}~\bibnamefont {Stoll}}, \bibinfo {author}
  {\bibfnamefont {G.}~\bibnamefont {Sch{\"u}tz}}, \bibinfo {author}
  {\bibfnamefont {G.~S.~D.}\ \bibnamefont {Beach}}, \ and\ \bibinfo {author}
  {\bibfnamefont {M.}~\bibnamefont {Kl{\"{a}}ui}},\ }\href@noop {} {\bibfield
  {journal} {\bibinfo  {journal} {Nature Nanotechnology}\ }\textbf {\bibinfo
  {volume} {13}},\ \bibinfo {pages} {170} (\bibinfo {year} {2017})}\BibitemShut
  {NoStop}%
\bibitem [{\citenamefont {Woo}\ \emph {et~al.}(2018)\citenamefont {Woo},
  \citenamefont {Song}, \citenamefont {Zhang}, \citenamefont {Ezawa},
  \citenamefont {Zhou}, \citenamefont {Liu}, \citenamefont {Weigand},
  \citenamefont {Finizio}, \citenamefont {Raabe}, \citenamefont {Park},
  \citenamefont {Lee}, \citenamefont {Choi}, \citenamefont {Min}, \citenamefont
  {Koo},\ and\ \citenamefont {Chang}}]{art:woo_skyrmion_nucleation}%
  \BibitemOpen
  \bibfield  {author} {\bibinfo {author} {\bibfnamefont {S.}~\bibnamefont
  {Woo}}, \bibinfo {author} {\bibfnamefont {K.~M.}\ \bibnamefont {Song}},
  \bibinfo {author} {\bibfnamefont {X.}~\bibnamefont {Zhang}}, \bibinfo
  {author} {\bibfnamefont {M.}~\bibnamefont {Ezawa}}, \bibinfo {author}
  {\bibfnamefont {Y.}~\bibnamefont {Zhou}}, \bibinfo {author} {\bibfnamefont
  {X.}~\bibnamefont {Liu}}, \bibinfo {author} {\bibfnamefont {M.}~\bibnamefont
  {Weigand}}, \bibinfo {author} {\bibfnamefont {S.}~\bibnamefont {Finizio}},
  \bibinfo {author} {\bibfnamefont {J.}~\bibnamefont {Raabe}}, \bibinfo
  {author} {\bibfnamefont {M.-C.}\ \bibnamefont {Park}}, \bibinfo {author}
  {\bibfnamefont {K.-Y.}\ \bibnamefont {Lee}}, \bibinfo {author} {\bibfnamefont
  {J.~W.}\ \bibnamefont {Choi}}, \bibinfo {author} {\bibfnamefont {B.-C.}\
  \bibnamefont {Min}}, \bibinfo {author} {\bibfnamefont {H.~C.}\ \bibnamefont
  {Koo}}, \ and\ \bibinfo {author} {\bibfnamefont {J.}~\bibnamefont {Chang}},\
  }\href@noop {} {\bibfield  {journal} {\bibinfo  {journal} {Nature
  Electronics}\ }\textbf {\bibinfo {volume} {1}},\ \bibinfo {pages} {288}
  (\bibinfo {year} {2018})}\BibitemShut {NoStop}%
\bibitem [{\citenamefont {Finizio}\ \emph {et~al.}(2019)\citenamefont
  {Finizio}, \citenamefont {Zeissler}, \citenamefont {Wintz}, \citenamefont
  {Mayr}, \citenamefont {We{\ss}els}, \citenamefont {Huxtable}, \citenamefont
  {Burnell}, \citenamefont {Marrows},\ and\ \citenamefont
  {Raabe}}]{art:finizio_SkyrmionNucleation}%
  \BibitemOpen
  \bibfield  {author} {\bibinfo {author} {\bibfnamefont {S.}~\bibnamefont
  {Finizio}}, \bibinfo {author} {\bibfnamefont {K.}~\bibnamefont {Zeissler}},
  \bibinfo {author} {\bibfnamefont {S.}~\bibnamefont {Wintz}}, \bibinfo
  {author} {\bibfnamefont {S.}~\bibnamefont {Mayr}}, \bibinfo {author}
  {\bibfnamefont {T.}~\bibnamefont {We{\ss}els}}, \bibinfo {author}
  {\bibfnamefont {A.~J.}\ \bibnamefont {Huxtable}}, \bibinfo {author}
  {\bibfnamefont {G.}~\bibnamefont {Burnell}}, \bibinfo {author} {\bibfnamefont
  {C.~H.}\ \bibnamefont {Marrows}}, \ and\ \bibinfo {author} {\bibfnamefont
  {J.}~\bibnamefont {Raabe}},\ }\href@noop {} {\bibfield  {journal} {\bibinfo
  {journal} {Nano Letters}\ }\textbf {\bibinfo {volume} {19}},\ \bibinfo
  {pages} {7246} (\bibinfo {year} {2019})}\BibitemShut {NoStop}%
\bibitem [{\citenamefont {B{\"u}ttner}\ \emph {et~al.}(2015)\citenamefont
  {B{\"u}ttner}, \citenamefont {Moutafis}, \citenamefont {Schneider},
  \citenamefont {Kr{\"u}ger}, \citenamefont {G{\"u}nther}, \citenamefont
  {Geilhufe}, \citenamefont {v.~Korff~Schmising}, \citenamefont {Mohanty},
  \citenamefont {Pfau}, \citenamefont {Schaffert}, \citenamefont {Bisig},
  \citenamefont {Foerster}, \citenamefont {Schulz}, \citenamefont {Vaz},
  \citenamefont {Franken}, \citenamefont {Swagten}, \citenamefont {Kl{\"a}ui},\
  and\ \citenamefont {Eisebitt}}]{art:buettner_gyration}%
  \BibitemOpen
  \bibfield  {author} {\bibinfo {author} {\bibfnamefont {F.}~\bibnamefont
  {B{\"u}ttner}}, \bibinfo {author} {\bibfnamefont {C.}~\bibnamefont
  {Moutafis}}, \bibinfo {author} {\bibfnamefont {M.}~\bibnamefont {Schneider}},
  \bibinfo {author} {\bibfnamefont {B.}~\bibnamefont {Kr{\"u}ger}}, \bibinfo
  {author} {\bibfnamefont {C.~M.}\ \bibnamefont {G{\"u}nther}}, \bibinfo
  {author} {\bibfnamefont {J.}~\bibnamefont {Geilhufe}}, \bibinfo {author}
  {\bibfnamefont {C.}~\bibnamefont {v.~Korff~Schmising}}, \bibinfo {author}
  {\bibfnamefont {J.}~\bibnamefont {Mohanty}}, \bibinfo {author} {\bibfnamefont
  {B.}~\bibnamefont {Pfau}}, \bibinfo {author} {\bibfnamefont {S.}~\bibnamefont
  {Schaffert}}, \bibinfo {author} {\bibfnamefont {A.}~\bibnamefont {Bisig}},
  \bibinfo {author} {\bibfnamefont {M.}~\bibnamefont {Foerster}}, \bibinfo
  {author} {\bibfnamefont {T.}~\bibnamefont {Schulz}}, \bibinfo {author}
  {\bibfnamefont {C.~A.~F.}\ \bibnamefont {Vaz}}, \bibinfo {author}
  {\bibfnamefont {J.~H.}\ \bibnamefont {Franken}}, \bibinfo {author}
  {\bibfnamefont {H.~J.~M.}\ \bibnamefont {Swagten}}, \bibinfo {author}
  {\bibfnamefont {M.}~\bibnamefont {Kl{\"a}ui}}, \ and\ \bibinfo {author}
  {\bibfnamefont {S.}~\bibnamefont {Eisebitt}},\ }\href@noop {} {\bibfield
  {journal} {\bibinfo  {journal} {Nature Physics}\ }\textbf {\bibinfo {volume}
  {11}},\ \bibinfo {pages} {225} (\bibinfo {year} {2015})}\BibitemShut
  {NoStop}%
\bibitem [{\citenamefont {Baumgartner}\ \emph {et~al.}(2017)\citenamefont
  {Baumgartner}, \citenamefont {Garello}, \citenamefont {Mendil}, \citenamefont
  {Avci}, \citenamefont {Grimaldi}, \citenamefont {Murer}, \citenamefont
  {Feng}, \citenamefont {Gabureac}, \citenamefont {Stamm}, \citenamefont
  {Ackermann}, \citenamefont {Finizio}, \citenamefont {Wintz}, \citenamefont
  {Raabe},\ and\ \citenamefont {Gambardella}}]{art:baumgartner_switching}%
  \BibitemOpen
  \bibfield  {author} {\bibinfo {author} {\bibfnamefont {M.}~\bibnamefont
  {Baumgartner}}, \bibinfo {author} {\bibfnamefont {K.}~\bibnamefont
  {Garello}}, \bibinfo {author} {\bibfnamefont {J.}~\bibnamefont {Mendil}},
  \bibinfo {author} {\bibfnamefont {C.~O.}\ \bibnamefont {Avci}}, \bibinfo
  {author} {\bibfnamefont {E.}~\bibnamefont {Grimaldi}}, \bibinfo {author}
  {\bibfnamefont {C.}~\bibnamefont {Murer}}, \bibinfo {author} {\bibfnamefont
  {J.}~\bibnamefont {Feng}}, \bibinfo {author} {\bibfnamefont {M.}~\bibnamefont
  {Gabureac}}, \bibinfo {author} {\bibfnamefont {C.}~\bibnamefont {Stamm}},
  \bibinfo {author} {\bibfnamefont {Y.}~\bibnamefont {Ackermann}}, \bibinfo
  {author} {\bibfnamefont {S.}~\bibnamefont {Finizio}}, \bibinfo {author}
  {\bibfnamefont {S.}~\bibnamefont {Wintz}}, \bibinfo {author} {\bibfnamefont
  {J.}~\bibnamefont {Raabe}}, \ and\ \bibinfo {author} {\bibfnamefont
  {P.}~\bibnamefont {Gambardella}},\ }\href@noop {} {\bibfield  {journal}
  {\bibinfo  {journal} {Nature Nanotechnology}\ }\textbf {\bibinfo {volume}
  {12}},\ \bibinfo {pages} {980} (\bibinfo {year} {2017})}\BibitemShut
  {NoStop}%
\bibitem [{\citenamefont {Fernandez-Pacheco}\ \emph {et~al.}(2017)\citenamefont
  {Fernandez-Pacheco}, \citenamefont {Streubel}, \citenamefont {Fruchart},
  \citenamefont {Hertel}, \citenamefont {Fischer},\ and\ \citenamefont
  {Cowburn}}]{art:fernandez_3DMagnetism}%
  \BibitemOpen
  \bibfield  {author} {\bibinfo {author} {\bibfnamefont {A.}~\bibnamefont
  {Fernandez-Pacheco}}, \bibinfo {author} {\bibfnamefont {R.}~\bibnamefont
  {Streubel}}, \bibinfo {author} {\bibfnamefont {O.}~\bibnamefont {Fruchart}},
  \bibinfo {author} {\bibfnamefont {R.}~\bibnamefont {Hertel}}, \bibinfo
  {author} {\bibfnamefont {P.}~\bibnamefont {Fischer}}, \ and\ \bibinfo
  {author} {\bibfnamefont {R.~P.}\ \bibnamefont {Cowburn}},\ }\href@noop {}
  {\bibfield  {journal} {\bibinfo  {journal} {Nature Communications}\ }\textbf
  {\bibinfo {volume} {8}},\ \bibinfo {pages} {15756} (\bibinfo {year}
  {2017})}\BibitemShut {NoStop}%
\bibitem [{\citenamefont {Fischer}\ \emph {et~al.}(2020)\citenamefont
  {Fischer}, \citenamefont {Sanz-Hernandez}, \citenamefont {Streubel},\ and\
  \citenamefont {Fernandez-Pacheco}}]{art:fisher_3DMagnetism}%
  \BibitemOpen
  \bibfield  {author} {\bibinfo {author} {\bibfnamefont {P.}~\bibnamefont
  {Fischer}}, \bibinfo {author} {\bibfnamefont {D.}~\bibnamefont
  {Sanz-Hernandez}}, \bibinfo {author} {\bibfnamefont {R.}~\bibnamefont
  {Streubel}}, \ and\ \bibinfo {author} {\bibfnamefont {A.}~\bibnamefont
  {Fernandez-Pacheco}},\ }\href@noop {} {\bibfield  {journal} {\bibinfo
  {journal} {APL Materials}\ }\textbf {\bibinfo {volume} {8}},\ \bibinfo
  {pages} {010701} (\bibinfo {year} {2020})}\BibitemShut {NoStop}%
\bibitem [{\citenamefont {Wartelle}\ \emph {et~al.}(2019)\citenamefont
  {Wartelle}, \citenamefont {Trapp}, \citenamefont {Stano}, \citenamefont
  {Thirion}, \citenamefont {Bochmann}, \citenamefont {Bachmann}, \citenamefont
  {Foerster}, \citenamefont {Aballe}, \citenamefont {Mentes}, \citenamefont
  {Locatelli}, \citenamefont {Sala}, \citenamefont {Cagnon}, \citenamefont
  {Toussaint},\ and\ \citenamefont {Fruchart}}]{art:wartelle_bloch_points}%
  \BibitemOpen
  \bibfield  {author} {\bibinfo {author} {\bibfnamefont {A.}~\bibnamefont
  {Wartelle}}, \bibinfo {author} {\bibfnamefont {B.}~\bibnamefont {Trapp}},
  \bibinfo {author} {\bibfnamefont {M.}~\bibnamefont {Stano}}, \bibinfo
  {author} {\bibfnamefont {C.}~\bibnamefont {Thirion}}, \bibinfo {author}
  {\bibfnamefont {S.}~\bibnamefont {Bochmann}}, \bibinfo {author}
  {\bibfnamefont {J.}~\bibnamefont {Bachmann}}, \bibinfo {author}
  {\bibfnamefont {M.}~\bibnamefont {Foerster}}, \bibinfo {author}
  {\bibfnamefont {L.}~\bibnamefont {Aballe}}, \bibinfo {author} {\bibfnamefont
  {T.~O.}\ \bibnamefont {Mentes}}, \bibinfo {author} {\bibfnamefont
  {A.}~\bibnamefont {Locatelli}}, \bibinfo {author} {\bibfnamefont
  {A.}~\bibnamefont {Sala}}, \bibinfo {author} {\bibfnamefont {L.}~\bibnamefont
  {Cagnon}}, \bibinfo {author} {\bibfnamefont {J.-C.}\ \bibnamefont
  {Toussaint}}, \ and\ \bibinfo {author} {\bibfnamefont {O.}~\bibnamefont
  {Fruchart}},\ }\href@noop {} {\bibfield  {journal} {\bibinfo  {journal}
  {Physical Review B}\ }\textbf {\bibinfo {volume} {99}},\ \bibinfo {pages}
  {024433} (\bibinfo {year} {2019})}\BibitemShut {NoStop}%
\bibitem [{\citenamefont {Donnelly}\ \emph {et~al.}(2017)\citenamefont
  {Donnelly}, \citenamefont {Guizar-Sicarios}, \citenamefont {Scagnoli},
  \citenamefont {Gliga}, \citenamefont {Holler}, \citenamefont {Raabe},\ and\
  \citenamefont {Heyderman}}]{art:donnelly_3DImaging}%
  \BibitemOpen
  \bibfield  {author} {\bibinfo {author} {\bibfnamefont {C.}~\bibnamefont
  {Donnelly}}, \bibinfo {author} {\bibfnamefont {M.}~\bibnamefont
  {Guizar-Sicarios}}, \bibinfo {author} {\bibfnamefont {V.}~\bibnamefont
  {Scagnoli}}, \bibinfo {author} {\bibfnamefont {S.}~\bibnamefont {Gliga}},
  \bibinfo {author} {\bibfnamefont {M.}~\bibnamefont {Holler}}, \bibinfo
  {author} {\bibfnamefont {J.}~\bibnamefont {Raabe}}, \ and\ \bibinfo {author}
  {\bibfnamefont {L.~J.}\ \bibnamefont {Heyderman}},\ }\href@noop {} {\bibfield
   {journal} {\bibinfo  {journal} {Nature}\ }\textbf {\bibinfo {volume}
  {547}},\ \bibinfo {pages} {328} (\bibinfo {year} {2017})}\BibitemShut
  {NoStop}%
\bibitem [{\citenamefont {Donnelly}\ \emph {et~al.}(2020)\citenamefont
  {Donnelly}, \citenamefont {Finizio}, \citenamefont {Gliga}, \citenamefont
  {Holler}, \citenamefont {Hrabec}, \citenamefont {Odstrcil}, \citenamefont
  {Mayr}, \citenamefont {Scagnoli}, \citenamefont {Heyderman}, \citenamefont
  {Guizar-Sicarios},\ and\ \citenamefont {Raabe}}]{art:donnelly_TRLamni}%
  \BibitemOpen
  \bibfield  {author} {\bibinfo {author} {\bibfnamefont {C.}~\bibnamefont
  {Donnelly}}, \bibinfo {author} {\bibfnamefont {S.}~\bibnamefont {Finizio}},
  \bibinfo {author} {\bibfnamefont {S.}~\bibnamefont {Gliga}}, \bibinfo
  {author} {\bibfnamefont {M.}~\bibnamefont {Holler}}, \bibinfo {author}
  {\bibfnamefont {A.}~\bibnamefont {Hrabec}}, \bibinfo {author} {\bibfnamefont
  {M.}~\bibnamefont {Odstrcil}}, \bibinfo {author} {\bibfnamefont
  {S.}~\bibnamefont {Mayr}}, \bibinfo {author} {\bibfnamefont {V.}~\bibnamefont
  {Scagnoli}}, \bibinfo {author} {\bibfnamefont {L.~J.}\ \bibnamefont
  {Heyderman}}, \bibinfo {author} {\bibfnamefont {M.}~\bibnamefont
  {Guizar-Sicarios}}, \ and\ \bibinfo {author} {\bibfnamefont {J.}~\bibnamefont
  {Raabe}},\ }\href@noop {} {\bibfield  {journal} {\bibinfo  {journal} {Nature
  Nanotechnology}\ }\textbf {\bibinfo {volume} {15}},\ \bibinfo {pages} {356}
  (\bibinfo {year} {2020})}\BibitemShut {NoStop}%
\bibitem [{\citenamefont {Witte}\ \emph {et~al.}(2020)\citenamefont {Witte},
  \citenamefont {Sp{\"a}th}, \citenamefont {Finizio}, \citenamefont {Donnelly},
  \citenamefont {Watts}, \citenamefont {Sarafimov}, \citenamefont {Odstrcil},
  \citenamefont {Guizar-Sicarios}, \citenamefont {Holler}, \citenamefont
  {Fink},\ and\ \citenamefont {Raabe}}]{art:witte_Lamni}%
  \BibitemOpen
  \bibfield  {author} {\bibinfo {author} {\bibfnamefont {K.}~\bibnamefont
  {Witte}}, \bibinfo {author} {\bibfnamefont {A.}~\bibnamefont {Sp{\"a}th}},
  \bibinfo {author} {\bibfnamefont {S.}~\bibnamefont {Finizio}}, \bibinfo
  {author} {\bibfnamefont {C.}~\bibnamefont {Donnelly}}, \bibinfo {author}
  {\bibfnamefont {B.}~\bibnamefont {Watts}}, \bibinfo {author} {\bibfnamefont
  {B.}~\bibnamefont {Sarafimov}}, \bibinfo {author} {\bibfnamefont
  {M.}~\bibnamefont {Odstrcil}}, \bibinfo {author} {\bibfnamefont
  {M.}~\bibnamefont {Guizar-Sicarios}}, \bibinfo {author} {\bibfnamefont
  {M.}~\bibnamefont {Holler}}, \bibinfo {author} {\bibfnamefont {R.~H.}\
  \bibnamefont {Fink}}, \ and\ \bibinfo {author} {\bibfnamefont
  {J.}~\bibnamefont {Raabe}},\ }\href@noop {} {\bibfield  {journal} {\bibinfo
  {journal} {Nano Letters}\ }\textbf {\bibinfo {volume} {20}},\ \bibinfo
  {pages} {1305} (\bibinfo {year} {2020})}\BibitemShut {NoStop}%
\bibitem [{\citenamefont {Holler}\ \emph {et~al.}(2019)\citenamefont {Holler},
  \citenamefont {Odstrcil}, \citenamefont {Guizar-Sicarios}, \citenamefont
  {Lebugle}, \citenamefont {M{\"u}ller}, \citenamefont {Finizio}, \citenamefont
  {Tinti}, \citenamefont {David}, \citenamefont {Zusman}, \citenamefont
  {Unglaub}, \citenamefont {Bunk}, \citenamefont {Raabe}, \citenamefont
  {Levi},\ and\ \citenamefont {Aeppli}}]{art:holler_lamni}%
  \BibitemOpen
  \bibfield  {author} {\bibinfo {author} {\bibfnamefont {M.}~\bibnamefont
  {Holler}}, \bibinfo {author} {\bibfnamefont {M.}~\bibnamefont {Odstrcil}},
  \bibinfo {author} {\bibfnamefont {M.}~\bibnamefont {Guizar-Sicarios}},
  \bibinfo {author} {\bibfnamefont {M.}~\bibnamefont {Lebugle}}, \bibinfo
  {author} {\bibfnamefont {E.}~\bibnamefont {M{\"u}ller}}, \bibinfo {author}
  {\bibfnamefont {S.}~\bibnamefont {Finizio}}, \bibinfo {author} {\bibfnamefont
  {G.}~\bibnamefont {Tinti}}, \bibinfo {author} {\bibfnamefont
  {C.}~\bibnamefont {David}}, \bibinfo {author} {\bibfnamefont
  {J.}~\bibnamefont {Zusman}}, \bibinfo {author} {\bibfnamefont
  {W.}~\bibnamefont {Unglaub}}, \bibinfo {author} {\bibfnamefont
  {O.}~\bibnamefont {Bunk}}, \bibinfo {author} {\bibfnamefont {J.}~\bibnamefont
  {Raabe}}, \bibinfo {author} {\bibfnamefont {A.~F.~J.}\ \bibnamefont {Levi}},
  \ and\ \bibinfo {author} {\bibfnamefont {G.}~\bibnamefont {Aeppli}},\
  }\href@noop {} {\bibfield  {journal} {\bibinfo  {journal} {Nature
  Electronics}\ }\textbf {\bibinfo {volume} {2}},\ \bibinfo {pages} {464}
  (\bibinfo {year} {2019})}\BibitemShut {NoStop}%
\bibitem [{\citenamefont {Puzic}\ \emph {et~al.}(2010)\citenamefont {Puzic},
  \citenamefont {Korhonen}, \citenamefont {Kalantari}, \citenamefont {Raabe},
  \citenamefont {Quitmann}, \citenamefont {J{\"u}llig}, \citenamefont {Bommer},
  \citenamefont {Goll}, \citenamefont {Sch{\"u}tz}, \citenamefont {Wintz},
  \citenamefont {Strache}, \citenamefont {K{\"o}rner}, \citenamefont {Marko},
  \citenamefont {Bunce},\ and\ \citenamefont {Fassbender}}]{art:puzic_TR_STXM}%
  \BibitemOpen
  \bibfield  {author} {\bibinfo {author} {\bibfnamefont {A.}~\bibnamefont
  {Puzic}}, \bibinfo {author} {\bibfnamefont {T.}~\bibnamefont {Korhonen}},
  \bibinfo {author} {\bibfnamefont {B.}~\bibnamefont {Kalantari}}, \bibinfo
  {author} {\bibfnamefont {J.}~\bibnamefont {Raabe}}, \bibinfo {author}
  {\bibfnamefont {C.}~\bibnamefont {Quitmann}}, \bibinfo {author}
  {\bibfnamefont {P.}~\bibnamefont {J{\"u}llig}}, \bibinfo {author}
  {\bibfnamefont {L.}~\bibnamefont {Bommer}}, \bibinfo {author} {\bibfnamefont
  {D.}~\bibnamefont {Goll}}, \bibinfo {author} {\bibfnamefont {G.}~\bibnamefont
  {Sch{\"u}tz}}, \bibinfo {author} {\bibfnamefont {S.}~\bibnamefont {Wintz}},
  \bibinfo {author} {\bibfnamefont {T.}~\bibnamefont {Strache}}, \bibinfo
  {author} {\bibfnamefont {M.}~\bibnamefont {K{\"o}rner}}, \bibinfo {author}
  {\bibfnamefont {D.}~\bibnamefont {Marko}}, \bibinfo {author} {\bibfnamefont
  {C.}~\bibnamefont {Bunce}}, \ and\ \bibinfo {author} {\bibfnamefont
  {J.}~\bibnamefont {Fassbender}},\ }\href@noop {} {\bibfield  {journal}
  {\bibinfo  {journal} {Synchrotron Radiation News}\ }\textbf {\bibinfo
  {volume} {23}},\ \bibinfo {pages} {26} (\bibinfo {year} {2010})}\BibitemShut
  {NoStop}%
\bibitem [{\citenamefont {Finizio}\ \emph {et~al.}(2020)\citenamefont
  {Finizio}, \citenamefont {Mayr},\ and\ \citenamefont
  {Raabe}}]{art:finizio_QuTAG}%
  \BibitemOpen
  \bibfield  {author} {\bibinfo {author} {\bibfnamefont {S.}~\bibnamefont
  {Finizio}}, \bibinfo {author} {\bibfnamefont {S.}~\bibnamefont {Mayr}}, \
  and\ \bibinfo {author} {\bibfnamefont {J.}~\bibnamefont {Raabe}},\
  }\href@noop {} {\bibfield  {journal} {\bibinfo  {journal} {Journal of
  Synchrotron Radiation}\ }\textbf {\bibinfo {volume} {27}},\ \bibinfo {pages}
  {1320} (\bibinfo {year} {2020})}\BibitemShut {NoStop}%
\bibitem [{\citenamefont {Donnelly}\ \emph {et~al.}(2018)\citenamefont
  {Donnelly}, \citenamefont {Gliga}, \citenamefont {Scagnoli}, \citenamefont
  {Holler}, \citenamefont {Raabe}, \citenamefont {Heyderman},\ and\
  \citenamefont {{Guizar-Sicarios}}}]{art:donnelly_3DReconstruction}%
  \BibitemOpen
  \bibfield  {author} {\bibinfo {author} {\bibfnamefont {C.}~\bibnamefont
  {Donnelly}}, \bibinfo {author} {\bibfnamefont {S.}~\bibnamefont {Gliga}},
  \bibinfo {author} {\bibfnamefont {V.}~\bibnamefont {Scagnoli}}, \bibinfo
  {author} {\bibfnamefont {M.}~\bibnamefont {Holler}}, \bibinfo {author}
  {\bibfnamefont {J.}~\bibnamefont {Raabe}}, \bibinfo {author} {\bibfnamefont
  {L.~J.}\ \bibnamefont {Heyderman}}, \ and\ \bibinfo {author} {\bibfnamefont
  {M.}~\bibnamefont {{Guizar-Sicarios}}},\ }\href@noop {} {\bibfield  {journal}
  {\bibinfo  {journal} {New Journal of Physics}\ }\textbf {\bibinfo {volume}
  {20}},\ \bibinfo {pages} {083009} (\bibinfo {year} {2018})}\BibitemShut
  {NoStop}%
\bibitem [{\citenamefont {Vansteenkiste}\ \emph {et~al.}(2014)\citenamefont
  {Vansteenkiste}, \citenamefont {Leliaert}, \citenamefont {Dvornik},
  \citenamefont {Helsen}, \citenamefont {{Garcia-Sanchez}},\ and\ \citenamefont
  {{Van Waeyenberge}}}]{art:mumax}%
  \BibitemOpen
  \bibfield  {author} {\bibinfo {author} {\bibfnamefont {A.}~\bibnamefont
  {Vansteenkiste}}, \bibinfo {author} {\bibfnamefont {J.}~\bibnamefont
  {Leliaert}}, \bibinfo {author} {\bibfnamefont {M.}~\bibnamefont {Dvornik}},
  \bibinfo {author} {\bibfnamefont {M.}~\bibnamefont {Helsen}}, \bibinfo
  {author} {\bibfnamefont {F.}~\bibnamefont {{Garcia-Sanchez}}}, \ and\
  \bibinfo {author} {\bibfnamefont {B.}~\bibnamefont {{Van Waeyenberge}}},\
  }\href@noop {} {\bibfield  {journal} {\bibinfo  {journal} {AIP Advances}\
  }\textbf {\bibinfo {volume} {4}},\ \bibinfo {pages} {107133} (\bibinfo {year}
  {2014})}\BibitemShut {NoStop}%
\bibitem [{\citenamefont {Makhfudz}\ \emph {et~al.}(2012)\citenamefont
  {Makhfudz}, \citenamefont {Kr{\"u}ger},\ and\ \citenamefont
  {Tchernyshyov}}]{art:makhfudz_gyration}%
  \BibitemOpen
  \bibfield  {author} {\bibinfo {author} {\bibfnamefont {I.}~\bibnamefont
  {Makhfudz}}, \bibinfo {author} {\bibfnamefont {B.}~\bibnamefont
  {Kr{\"u}ger}}, \ and\ \bibinfo {author} {\bibfnamefont {O.}~\bibnamefont
  {Tchernyshyov}},\ }\href@noop {} {\bibfield  {journal} {\bibinfo  {journal}
  {Physical Review Letters}\ }\textbf {\bibinfo {volume} {109}},\ \bibinfo
  {pages} {217201} (\bibinfo {year} {2012})}\BibitemShut {NoStop}%
\bibitem [{\citenamefont {Finizio}\ \emph {et~al.}(2021)\citenamefont
  {Finizio}, \citenamefont {Watts},\ and\ \citenamefont
  {Raabe}}]{art:finizio_TR_STXM_simulations}%
  \BibitemOpen
  \bibfield  {author} {\bibinfo {author} {\bibfnamefont {S.}~\bibnamefont
  {Finizio}}, \bibinfo {author} {\bibfnamefont {B.}~\bibnamefont {Watts}}, \
  and\ \bibinfo {author} {\bibfnamefont {J.}~\bibnamefont {Raabe}},\
  }\href@noop {} {\bibfield  {journal} {\bibinfo  {journal} {Journal of
  Synchrotron Radiation}\ }\textbf {\bibinfo {volume} {28}},\ \bibinfo {pages}
  {1146} (\bibinfo {year} {2021})}\BibitemShut {NoStop}%
\bibitem [{\citenamefont {Raabe}\ \emph {et~al.}(2008)\citenamefont {Raabe},
  \citenamefont {Tzvetkov}, \citenamefont {Flechsig}, \citenamefont {{B\"oge}},
  \citenamefont {Jaggi}, \citenamefont {Sarafimov}, \citenamefont {Vernooij},
  \citenamefont {Huthwelker}, \citenamefont {Ade}, \citenamefont {Kilcoyne},
  \citenamefont {Tyliszczak}, \citenamefont {Fink},\ and\ \citenamefont
  {Quitmann}}]{art:pollux}%
  \BibitemOpen
  \bibfield  {author} {\bibinfo {author} {\bibfnamefont {J.}~\bibnamefont
  {Raabe}}, \bibinfo {author} {\bibfnamefont {G.}~\bibnamefont {Tzvetkov}},
  \bibinfo {author} {\bibfnamefont {U.}~\bibnamefont {Flechsig}}, \bibinfo
  {author} {\bibfnamefont {M.}~\bibnamefont {{B\"oge}}}, \bibinfo {author}
  {\bibfnamefont {A.}~\bibnamefont {Jaggi}}, \bibinfo {author} {\bibfnamefont
  {B.}~\bibnamefont {Sarafimov}}, \bibinfo {author} {\bibfnamefont {M.~G.~C.}\
  \bibnamefont {Vernooij}}, \bibinfo {author} {\bibfnamefont {T.}~\bibnamefont
  {Huthwelker}}, \bibinfo {author} {\bibfnamefont {H.}~\bibnamefont {Ade}},
  \bibinfo {author} {\bibfnamefont {D.}~\bibnamefont {Kilcoyne}}, \bibinfo
  {author} {\bibfnamefont {T.}~\bibnamefont {Tyliszczak}}, \bibinfo {author}
  {\bibfnamefont {R.~H.}\ \bibnamefont {Fink}}, \ and\ \bibinfo {author}
  {\bibfnamefont {C.}~\bibnamefont {Quitmann}},\ }\href@noop {} {\bibfield
  {journal} {\bibinfo  {journal} {Review of Scientific Instruments}\ }\textbf
  {\bibinfo {volume} {79}},\ \bibinfo {pages} {113704} (\bibinfo {year}
  {2008})}\BibitemShut {NoStop}%
\bibitem [{\citenamefont {Ahrens}\ \emph {et~al.}(2005)\citenamefont {Ahrens},
  \citenamefont {Geveci},\ and\ \citenamefont {Law}}]{book:paraview}%
  \BibitemOpen
  \bibfield  {author} {\bibinfo {author} {\bibfnamefont {J.}~\bibnamefont
  {Ahrens}}, \bibinfo {author} {\bibfnamefont {B.}~\bibnamefont {Geveci}}, \
  and\ \bibinfo {author} {\bibfnamefont {C.}~\bibnamefont {Law}},\ }\href@noop
  {} {\emph {\bibinfo {title} {ParaView: an end-user tool for large data
  visualization}}}\ (\bibinfo  {publisher} {Elsevier},\ \bibinfo {year}
  {2005})\BibitemShut {NoStop}%
\end{thebibliography}
\end{document}